\documentclass[aps,prd,eqsecnum,superscriptaddress,twocolumn,nofootinbib,floatfix,preprintnumbers]{revtex4-2}

\usepackage{url,hyperref}
\usepackage{aas_macros}
\usepackage{multirow}
\usepackage{graphicx}
\usepackage{footnote}
\usepackage{natbib}
\usepackage[fleqn]{amsmath}
\usepackage{amsmath}
\usepackage{color}
\usepackage[title,toc,page]{appendix}
\usepackage{graphicx}
\usepackage{amssymb}
\usepackage{amsmath}
\usepackage{xcolor}
\usepackage{multirow}
\usepackage{graphicx}
\usepackage{subcaption}
\usepackage{hyphenat}
\usepackage{ORCIDinREVTeX}

\begin{document}

\newcommand*{\TTU}{Department of Physics and Astronomy, Texas Tech University, Lubbock, TX 79409-1051, USA}\affiliation{\TTU}
\newcommand*{\RICE}{Department of Physics and Astronomy, Rice University, Main Street,
Houston, TX 77005, USA}\affiliation{\RICE}
\newcommand*{\JHU}{William H. Miller III Department of Physics and Astronomy, Johns Hopkins University, Baltimore, MD 21218, USA}\affiliation{\JHU}
\newcommand*{\URI}{Department of Physics, University of Rhode Island, Kingston, Rhode Island 02881 (USA)}\affiliation{\URI}

\title{Searches for Post-Merger Gravitational Waves with CoCoA: Sensitivity Projections Across Large Template Banks for Current and Next-Generation Detectors}

\author{Tanazza Khanam}\email{tk73@rice.edu}\orcid{0000-0002-2483-5569}\affiliation{\TTU}\affiliation{\RICE}
\author{Alessandra Corsi}\email{acorsi2@jh.edu}\orcid{0000-0001-8104-3536}\affiliation{\JHU}
\author{Robert Coyne}\orcid{0000-0002-5243-5917}\affiliation{\URI}
\author{Michael St. Pierre}\orcid{0009-0008-7969-5381}\affiliation{\URI}
\date[\relax]{compiled \today }

\begin{abstract}
 The multi-messenger detection of the binary neutron star (NS) merger GW170817 has revolutionized the field of gravitational wave (GW) astronomy. However, several important questions remain to be answered.  One of these is the nature of the compact remnant leftover by GW170817 (short- or long-lived NS versus black hole). A key goal going forward is to understand the diversity of NS-NS merger remnants, and how such diversity maps onto their viability as gamma-ray burst (GRB) central engines. Here, we present a study aimed at assessing the sensitivity of triggered searches for intermediate-duration, post-merger GWs powered by long-lived GRB remnants using networks of current and future ground-based GW detectors and the Cross-Correlation Algorithm (CoCoA). 
 We develop a Python-based framework to efficiently estimate CoCoA distance horizons for a broad range of post-merger secular bar-mode waveforms and for different GW detector networks. This framework can be used to identify the most promising regions of parameter space in which to concentrate search efforts, helping to design future search strategies to optimally balance search sensitivity and related parameter space gridding schema against computational cost.
\end{abstract}

\flushbottom 
\maketitle

\thispagestyle{empty}

\section{Introduction}\label{s1}
On 2017 August 17, the LIGO and Virgo ground-based gravitational wave (GW) detectors identified their first binary neutron star (NS) merger\cite{2017PhRvL.119p1101A}, an event called GW170817. This event was, and as of today remains, the first and only NS-NS system for which an electromagnetic (EM) counterpart was observed in practically all bands of the spectrum (see \citep{2017ApJ...848L..12A} and references therein). The associated detection of a gamma-ray burst (GRB; \citep[][]{2017ApJ...848L..13A}), followed by a UV-optical-IR kilonova \citep{2017ApJ...848L..19C,2017ApJ...848L..17C,2017Sci...358.1570D,2017Sci...358.1565E,2017Sci...358.1559K,2017ApJ...848L..18N,2017Natur.551...67P,2017Natur.551...75S,2017ApJ...848L..16S,2017ApJ...848L..27T,2017ApJ...848L..24V,2017ApJ...851L..21V}, and a radio-to-X-ray afterglow \citep{2017Sci...358.1579H,2017ApJ...848L..21A,2017ApJ...848L..25H,2017ApJ...848L..20M,2017Natur.551...71T,2018Natur.554..207M,2018ApJ...856L..18M,2018Natur.561..355M}, were critical to drawing a detailed picture of the GW170817 merger.

Although the multi-messenger observations of GW170817 greatly improved our understanding of short GRBs, their relation to compact binary mergers, and the physics of their merger ejecta, several important questions remain open (see, e.g., \citep{2024FrASS..1186748C} and references therein). One of these is the nature of the merger remnant, which has motivated searches for post-merger, intermediate-to-long duration GW signals in GW170817 \citep{2017ApJ...851L..16A,2019ApJ...875..160A} and, more generally, in broader all-sky searches \citep{2018CQGra..35f5009A,2021PhRvD.104j2001A,2025arXiv250712282T}. Depending on the Equation of State (EoS) of nuclear matter, and hence on the maximum mass of a cold, non-rotating NS, a NS-NS merger could result in a promptly-formed black hole (BH), a Hyper-Massive NS, a Supra-Massive NS, or a stable NS (e.g., \citep{2017ApJ...850L..19M,Metzger2019LRR, Sarin_Lasky_2021} and references therein). Understanding how the diversity of remnant outcomes maps onto the GW-probed properties (e.g., total mass and mass ratio) of their progenitors, and how it impacts the physics of the merger ejecta and of potential GRB central engines (as probed by EM observations; \citep{Nousek_GRB, Zhang_2006, starling_2008, Bernardini_2012, Gompertz_2013, Rowlinson_2013, Yi_2014, Ravi_Lasky_2014}), is a key goal going forward. A particularly compelling scenario is one in which post-merger GW searches yield a direct signature of the merger remnant, complementing EM observations that may point to the presence of a long-lived remnant, such as a GRB X-ray plateau \citep{CM09,Gompertz2014}, X-ray flares \citep{Shan2024}, a bluer and brighter kilonova \citep{2017ApJ...850L..19M,2019ApJ...880L..15M,2022MNRAS.516.4949S}, or a late-time radio flare \citep{2011Natur.478...82N,2022MNRAS.516.4949S,Li2024}.

Among the proposed techniques for post-merger GW searches in LIGO data (e.g., \citep{2018CQGra..35f5009A,2019PhRvD..99j4033A,2021PhRvD.104j2001A,2025arXiv250712282T} and references therein) is the Cross-Correlation Algorithm (hereafter, CoCoA; \citep{Coyne2016}). This method targets quasi-periodic intermediate-duration GW signals ($10^2$--$10^4$\,s), such as those that may be observed in secularly-unstable newly-born NSs \citep{CM09,Lai_shapiro_1995}.  The proof of concept of CoCoA was first presented in \citeauthor{Coyne2016} (2016) where sensitivity estimates were provided for single-waveform searches. Then, in \citeauthor{Sowell2019} (2019) the method was generalized to enable searches on multiple waveforms (template bank), and sensitivity results were  presented for a search on real LIGO data using a small template bank. 

Anticipating the need for an increasing number of post-merger searches over  large template banks in data from ground-based GW detectors of improved sensitivity \citep{PostO5report} and next generation \citep{2023_cosmic_explorer_white_paper,2023JCAP...07..068B}, here we extend the work of \citeauthor{Coyne2016} and \citeauthor{Sowell2019} by developing a Python-based algorithm to analytically estimate CoCoA distance horizons for various detector networks over large template banks, and help quantify the effect of specific parameter-space gridding schema on the sensitivity of the search. We envision the algorithm developed as part of this work to become a useful pre-processing tool aimed at reducing the computational cost of CoCoA searches performed using real detector data. Specifically, the results of this work can be used to identify, for a given compact
binary merger system, the most promising region(s) of the parameter space where a search could
focus, and a reasonable step size for the envisioned parameter space grid.

This paper is organized as follows. In Section \ref{s2A}, we concisely explain the cross-correlation statistics on which CoCoA is based. In Section \ref{s2B} we summarize the properties of the class of bar-mode post-merger GW waveforms that we use as test bed for this work. In Section \ref{s2C} we discuss various detector network configurations of interest. In Sections \ref{s3} we describe our method for parameter space exploration. In Section \ref{s4} we present test-case results for GW170817-like post-merger searches with CoCoA. In Section \ref{s4C} we compare our results with previously published ones, and in Section \ref{s5} we discuss and conclude. 

\section{Background}\label{s2}
\subsection{CoCoA in a nutshell}\label{s2A}
CoCoA, originally developed by \citet{Coyne2016}, is a template-bank-based search algorithm that adapts the cross-correlation technique by \citet{Dhurandhar2009} to the search for GW signals of intermediate duration, i.e.,  much shorter than continuous GWs (e.g.,\cite{2019ApJ...875..160A,2022PhRvD.106j2008A})
but longer than typical GW burst searches (e.g., \cite{2021PhRvD.104l2004A, Marek_all_sky_short_2022,2025arXiv250712374L}).
In what follows, we refer to these as signals of intermediate duration. 

CoCoA searches are designed for GW signals $h(t)$ that are
quasi-periodic, i.e.,  signals that can be considered
monochromatic over small time intervals. For quasi-periodic signals, it is reasonable to assume that the time-frequency evolution can be modeled with sufficient physical accuracy over a time interval $T_{\rm coh}$, less than or equal to the total
observation time $T_{\rm obs}$ over which the signal is expected to last. Since the signal is quasi-periodic, we can also define a baseline $\Delta T_{\rm SFT} \le T_{\rm coh}$ such that, when performing a short Fourier Transform (SFT) of the signal within the baseline, all of the signal power is concentrated in a single SFT bin. More specifically, around each time $T_I$ we can approximate the signal received by the detector from $T_I-\Delta T_{\rm SFT}$ to 
$T_I + \Delta T_{\rm SFT}$, as:
\begin{eqnarray}
\nonumber h(t) \approx h_0(T_I){\cal A}_+F_+\cos(\phi(T_I)+2\pi f(T_I)(t-T_I))\\ + h_0(T_I){\cal A}_{\times}F_{\times}\sin(\phi(T_I)+2\pi f(T_I)(t-T_I)),~~~
\label{eq:amplitude-at-det}
\end{eqnarray}
where ${\mathcal A_+}$ and ${\mathcal A}_{\times}$ are amplitude factors dependent on the
physical system’s inclination angle $\iota$:
\begin{eqnarray}
{\mathcal A}_{+}=\frac{1+\cos^2{\iota}}{2}\\
{\mathcal A}_{\times}=\cos{\iota}.
\end{eqnarray}
While not strictly necessary, hereafter we assume for simplicity that $\iota=0$ (as in the case of a post-merger search associated with an on-axis short GRB). Also, in the above Equation, $F_+$ and $F_{\times}$ are the antenna factors that quantify the detector’s sensitivity to each GW polarization state. Here, we assume that $F_{+}$ and $F_{\times}$ are constant with time, given the duration of the signals we are interested in (see also \citep{Coyne2016}).

With CoCoA, a cross-correlation statistic is calculated from SFTs of detector's data along the model time-frequency tracks that are included in the search template bank. The specific way in which the CoCoA detection statistic is calculated results in a trade-off between robustness against model uncertainties and search sensitivity, where improved sensitivity comes at the expense of increased computational cost. In the so-called stochastic limit of CoCoA, only SFTs pairs from different detectors at the same time (after correcting for the time-of-flight in case of non co-located detectors) are correlated. This choice minimizes computational cost and maximizes robustness against signal uncertainties, at the expense of sensitivity. 
In the matched-filter limit of CoCoA, all possible SFT pairs (including self-pairs) are considered. This limit maximizes sensitivity at the expense of robustness and computing time. 
Finally, in the semi-coherent regime of CoCoA, the total observation time $T_{\rm obs}$ is broken up into $N_{\rm coh} = T_{\rm obs} / T_{\rm coh}$  coherent segments, each of duration $T_{\rm coh}$. The coherence time is defined as the length of time wherein the signal is expected to maintain phase coherence (and therefore good agreement) with the model predictions. All possible SFT-pairs within each coherent time segment are cross-correlated, and the results for each coherent time segment are then combined incoherently. The CoCoA semi-coherent approach can be regarded as the incoherent sum of $N_{\rm coh}$ matched-filter searches carried out over $N_{\rm coh}$ time segments each of duration $T_{\rm coh}$.

\subsection{Targeted GW signals}\label{s2B}
Among the models that predict efficient GW emission from a post-merger remnant are those that foresee the formation of a magnetar (a highly-magnetized NS) after the merger. Deformations of the magnetar could be induced by magnetic field distortions (e.g., \cite{Bonazzola_1996A&A}), unstable r-modes (e.g., \cite{Andersson_1998A}), and unstable bar modes (e.g., \cite{Lai_shapiro_1995}). 

The intermediate-duration GW waveform models that have been traditionally used in testing CoCoA are those of magnetars that are secularly unstable for bar-mode deformations. These are associated with NSs for which the kinetic energy to gravitational binding energy ratio ($\beta = T/|W|$) falls within the range of $0.14 < \beta < 0.27$ \citep{Lai_shapiro_1995}. Secularly unstable magnetars have also been proposed as possible central engines of GRBs showing plateaus in their X-ray afterglow light curves \citep{CM09}. 
For secularly unstable magnetars, the NS spin-down is due to energy losses associated with both the magnetic field and the GW emission, and can be quantified as follows (see Equation (11) in \citep{CM09}):
\begin{equation}
\frac{dE}{dt} = \frac{dE_{\rm{EM}}}{dt} + \frac{dE_{\rm{GW}}}{dt} = -\frac{B^2R^6\Omega_{\text{eff}}^4}{6c^3} - \frac{32GI^2\epsilon^2\Omega^6}{5c^5},
\label{eq:NS-SpinDown}
\end{equation}
where $dE_{\rm GW}/dt$ represents the GW energy losses; $dE_{\rm EM}/dt$ accounts for losses due to magnetic dipole radiation; $B$ is the magnetic dipole field strength at the poles; $R$ is the geometric mean of the principal axes of the deformed NS; $\Omega$ is the pattern angular frequency of the ellipsoidal-shaped NS; $\Omega_{\rm{eff}}$ is an effective angular frequency which includes both the ellipsoidal pattern speed and the effects of the internal motions of fluid particles; $\epsilon=(a^2_1-a^2_2)/(a^2_1+a^2_2)$ is the ellipticity (with $a_1$ and $a_2$ as the principal axes of the ellipsoidal figure in the equatorial plane); and $I$ is the moment of inertia with respect to the NS rotation axis. 

The GW losses from the spin-down law above result in a quasi-periodic GW signal of the form in Equation \ref{eq:amplitude-at-det} with frequency $ f(t) = \Omega(t)/\pi$ and amplitude (see Equation (14) in \citep{CM09}):
\begin{equation}
h_0(t) = \frac{4G\Omega(t)^2}{c^{4}d}I(t)\epsilon(t),
\label{eq:GW_strain}
\end{equation}
where $d$ is the distance to the source, $c$ is the speed of light, and $G$ is the gravitational constant.

In a previous paper, \citeauthor{Coyne2016} used two bar-mode waveforms (CM09long and CM09short) to derive CoCoA sensitivity estimates, while \citeauthor{Sowell2019}  used six bar-mode waveforms (Bar 1-6; see Table-1 of \citep{Sowell2019}) that were the focus of a GW170817 post-merger search on LIGO data \cite{2017ApJ...851L..16A}. Here, 
we build a larger template bank of GW waveforms suited for a post-merger remnant search similar to that presented in \citep{2017ApJ...851L..16A}. Namely, we set $T/|W|=0.2$, and mass $M$ of the NS remnant to 2.6\,$M_{\odot}$. The last is somewhere between the lower bounds of GW170817 estimated remnant mass 2.73$M_{\odot}$ and that of the total mass range of other known binary systems 2.57$M_{\odot}$(see \citep{2017ApJ...851L..16A} and references therein). We also assume the same range of EoS-motivated radii as in \citep{2017ApJ...851L..16A}, namely, $R=12-14$\,km, but differently from \citep{2017ApJ...851L..16A} we span this range with a finely spaced grid. Following \citep{2017ApJ...851L..16A}, we explore a range of magnetic field strengths $B=10^{13}-5\times$ 10$^{14}$\,G, which is suitable for magnetars that may be formed in NS-NS mergers and constitute central engines of associated short GRBs. Extremes beyond 10$^{15}$\,G lead to magnetic-dipole dominated spin-down of the NS and hence sub-dominant energy emission in GWs. Values below 10$^{13}$\,G are less likely in young magnetars given the post-merger
remnant dynamics that wind up strong fields (see e.g., \citep{2022PhRvD.106b3013P,2022ApJ...926L..31A} and references therein). 

\begin{figure}
    \centering
    \includegraphics[width=0.5\textwidth]{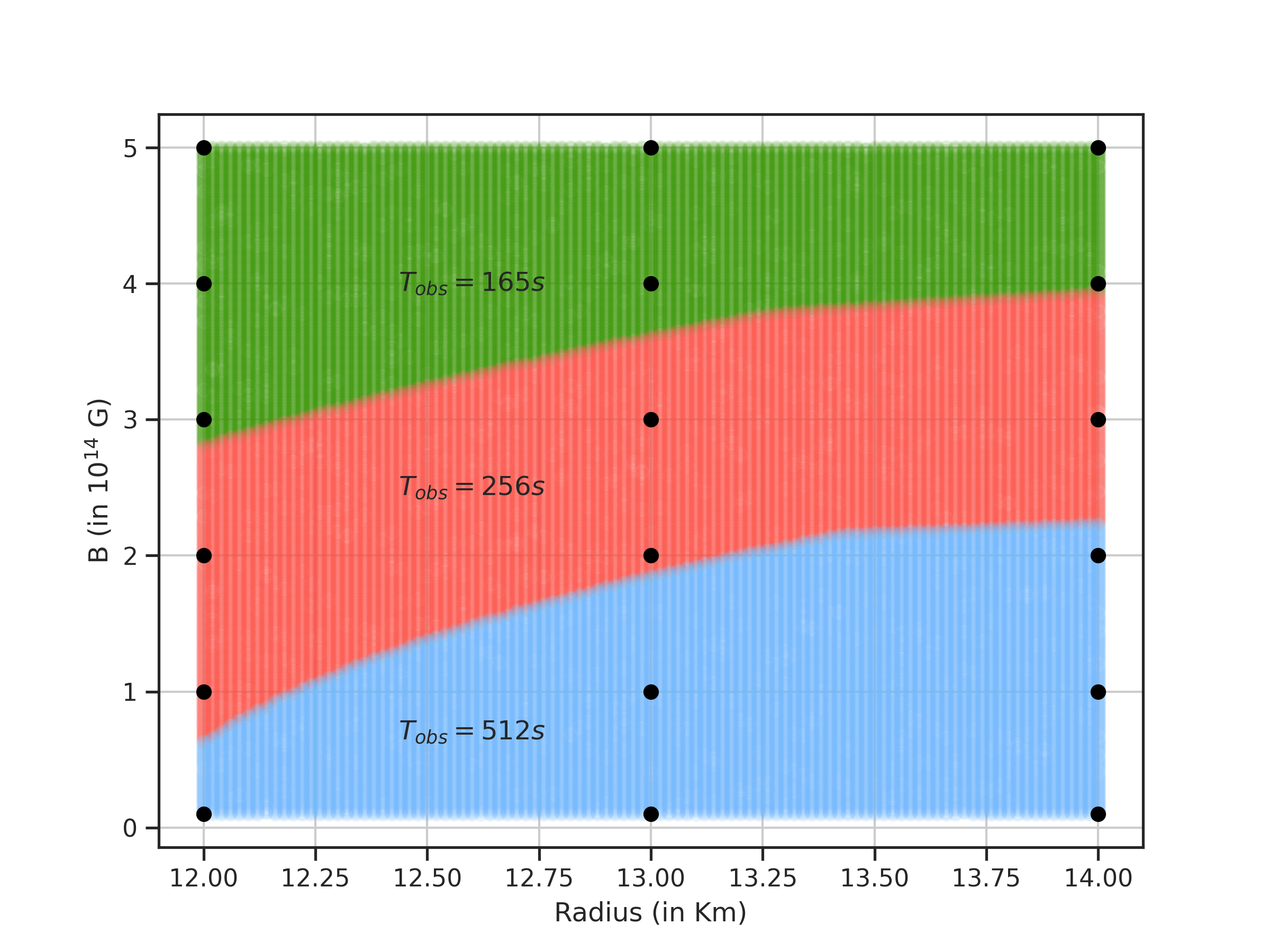}
   \caption{The parameter space in dipolar magnetic field strength ($B$) and magnetar radius ($R$), of a GW170817-like post-merger search for secular bar-mode GW signals. This parameter space can be broadly divided into different observation times ($T_{\text{obs}}$), ranging from the fastest evolving waveforms (green) to the slowest (blue). }
    \label{fig:parameter-space-black-dots}
\end{figure}

As described in \citep{2017ApJ...851L..16A}, because we do not
know the ultimate fate of the bar-shaped remnant and how long it can survive before collapsing to a BH, we consider the model waveform time-frequency evolution up to a time when the
luminosity emitted in GWs is 1\% of the peak value, which is
sufficient to capture the bulk of the energy emitted in GWs. As evident from Table 2 in \citep{2017ApJ...851L..16A}, the so-defined waveforms' durations can vary quite a bit. Hence, we group together waveforms of similar duration and define three major regions in our parameter space, with $T_{\rm obs}$ ranging between 165\,s and 512\,s (see Figure \ref{fig:parameter-space-black-dots}). This choice is purely phenomenological, and simply aimed at grouping together similar number of waveforms across a small number of observing time intervals such that waveforms in a given $T_{\rm obs}$ family can be reasonably analyzed with that choice of $T_{\rm obs}$ without substantial loss of signal power (which can occur if the adopted $T_{\rm obs}$ is too short compared to the timescale during which the bulk of the GW energy is emitted) and without introducing unneeded computational burden (which can occur if the adopted $T_{\rm obs}$ is too long compared to the timescale over which the bulk of the GW energy is emitted).

It is worth noting that the waveforms used here are representative of only one of the post-merger scenarios proposed in the literature. CoCoA itself can in principle be run on any intermediate-duration quasi-monochromatic waveform (see e.g.  \citep{2021PhRvD.104j2001A,2025arXiv250712282T} and references therein for more examples).

\subsection{Detector networks}\label{s2C}

\begin{figure}[t]
\includegraphics[width=\columnwidth]{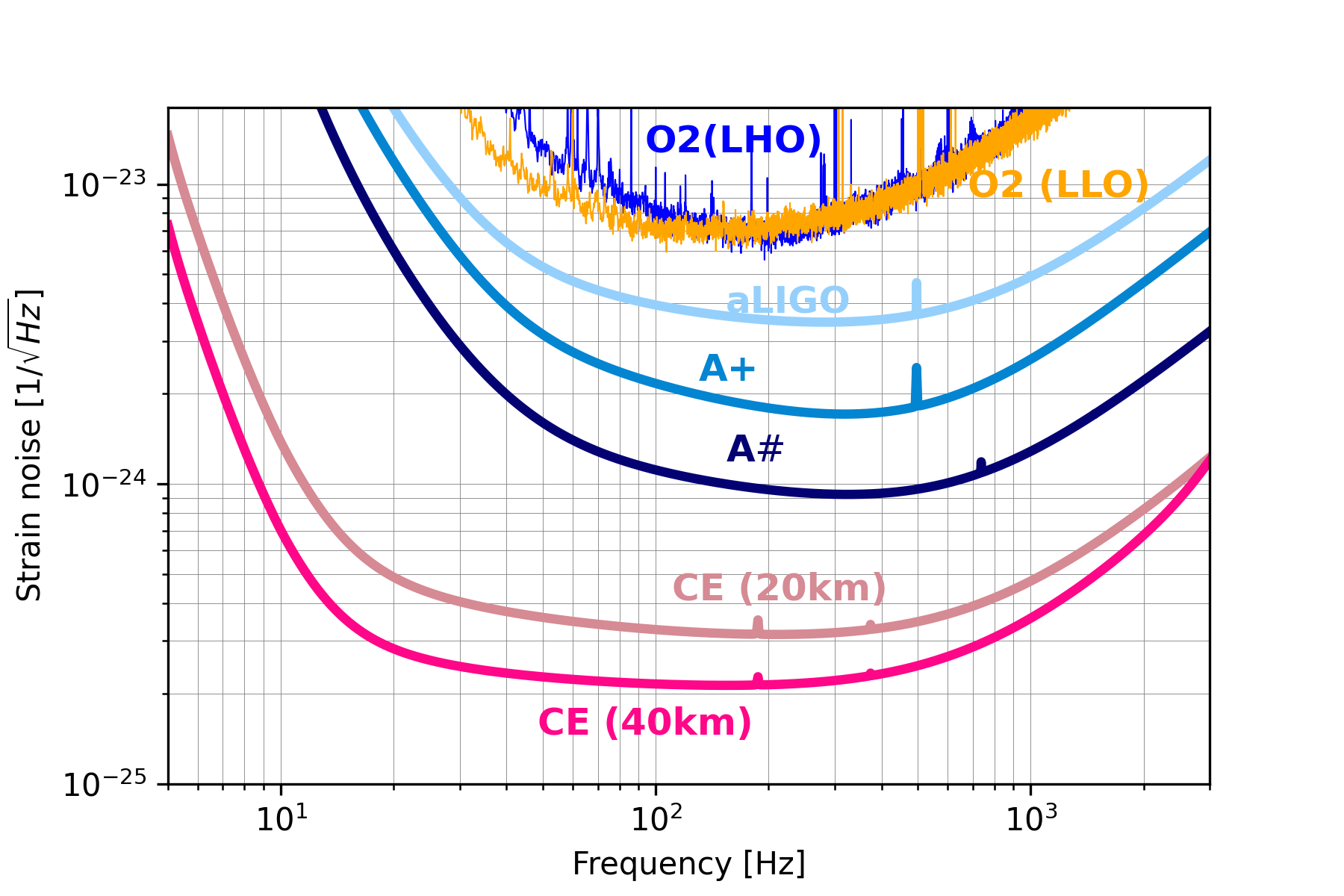}
    \caption{ Strain noise sensitivity curves of ground-based GW detectors. The O2 sensitivity curves are derived from real data, the others are expected sensitivities. See the caption of Table \ref{tab:detector-network} for details.} 
    \label{fig:Det_Net}
\end{figure}

In this work, we extend previous results also by estimating the performance of post-merger GW searches with CoCoA for different detector networks containing two detectors of comparable sensitivities. We note that, as discussed in \citet{Coyne2016}, CoCoA analyses can naturally incorporate either a single detector or an arbitrary number of detectors, with no need to modify the statistic. However, hereafter we consider a variety of two-detector networks to more closely compare our results with other two-detector searches (e.g., \citep{Sowell2019,2017ApJ...851L..16A,2025arXiv250712282T}) and to emphasize relative improvements related to detector sensitivity rather than to the number of detectors in the network. Figure~\ref{fig:Det_Net} shows sensitivity curves ranging from the LIGO second observing run (O2, the run during which GW170817 was detected) to Cosmic Explorer (CE) \citep{2023_cosmic_explorer_white_paper,ngGWreport}, a next-generation GW detector likely to operate on timescales similar to those envisioned for the  Einstein Telescope (ET) \citep{2023JCAP...07..068B}.
Table \ref{tab:detector-network} shows also various combinations of these detectors.  
Using these detector networks we aim to understand how future CoCoA searches could improve the distance reach at which we can probe post-merger scenarios.  We specifically limit our study to networks of two detectors only so as to be able to compare results with previous analyses \citep{Coyne2016,Sowell2019}.
\begin{center}
\begin{table}
\begin{tabular}{cccc}
\hline
\# of Detectors  & Network & Sensitivity & Detectors in Network \\
\hline
$2\times$2G & O2HL & O2& LLO \& LHO \\
$2\times$2G & O4HL &aLIGO & LLO \& LHO \\
$2\times$2G & O5HL &A+& LLO \& LHO\\
$2\times$2G & A$^\#$HL & A$^\#$& LLO \& LHO\\
$2\times$3G & CE4020 &CE& CE-40km \& CE-20km\\
\hline
\hline
\end{tabular}
 \caption{GW detector networks considered in this study. The O2HL network contains LIGO-Hanford (LHO) and LIGO-Livingston (LLO) at O2 sensitivity for a direct comparison with results presented in \citep{Sowell2019}. The O4HL network contains LLO and LHO at O4 sensitivity \citep{2018LRR....21....3A} for a direct comparison with projections presented in \citep{Coyne2016}. O5HL refers to LLO and LHO at the planned O5 sensitivity (also referred to as A+ \citep{2018LRR....21....3A}). A$^\#$HL refers to LHO and LLO at the envisioned post-O5 (also referred to as A$^\#$ \citep{PostO5report}) sensitivity. CE4020 is the network containing two next-generation GW detectors, Cosmic Explorer (CE) 40\,km and 20\,km \citep{2023_cosmic_explorer_white_paper,ngGWreport}. See Figure \ref{fig:Det_Net} for corresponding sensitivity curves.}
    \label{tab:detector-network}
\end{table}
\end{center}

\section{Methods}\label{s3}
As discussed in Section \ref{s1}, with the improvement in sensitivity of current and future GW detectors, the number of NS-NS merger detections per year is projected to greatly increase \citep{2023arXiv230710421G,2025arXiv250622835M}. This increase may pose a significant challenge for CoCoA post-merger searches that, for each merger, need the computing time required to cover what in general can be a rather large template bank, even when constraints derived from the in-spiral GW signal and/or from EM counterpart emission can reduce the parameter space of the search (see \citep{Sowell2019} for discussion). 

In the case of a GW170817-like post merger search within the secularly unstable magnetar model (see Section \ref{s2B}), one can explore a reduced 2-dimensional parameter space in $B$ and $R$ (see Section \ref{s2B}). However, even in this case, the number of templates required for a realistic search is {\cal O}($10^4-10^5$)   when using a conservative parameter space gridding scheme \citep{Coyne2016}, making the analysis computationally demanding \citep{Sowell2019}. 

Motivated by the above considerations, here we develop a Python-based code to estimate the expected horizon distances for CoCoA single-waveform post-merger searches of secular bar-mode GWs spanning the $(B, R)$ ranges described in Section \ref{s2B}, assuming Gaussian detector noise. These estimates extend the work presented in \citep{Coyne2016} by incorporating various detector networks (see Section \ref{s2C} and Table \ref{tab:detector-network}), and the ability to estimate horizon distances in all CoCoA limits (stochastic, semi-coherent, and matched filtering; see Section \ref{s2A}). While real (multi-waveform) searches will be affected by a trial factor that depends on the total number of templates included in the bank (see \citep{Sowell2019} for details), single-waveform search sensitivity estimates across large template banks can bracket the sensitivity achievable by any multi-waveform search, establishing an upper bound on their distance reach and helping identify, quickly and with little computational cost, regions of the parameter space that may not be worth exploring (due to reduced sensitivity of the search) if computational cost is high and computing resources are limited.

Following \citep{Coyne2016}, we assume a white-Gaussian noise power spectral density (PSD; see also Figure 2 of \citeauthor{Coyne2016}). For each waveform, assuming a post-merger NS located at a certain distance $d$, we can compute the root mean square (rms) amplitude at the detector as (see Equation 4.24 of \citep{Coyne2016}):
\begin{equation}
    h_{\rm rms} = \sqrt{\frac{\sum_{I} h_0^2(T_I)}{N_{\rm SFT}}},
    \label{eq:h_input}
\end{equation}
where $I$ is the SFT index; $N_{\rm SFT}=T_{\rm obs}/\Delta T_{\rm SFT}$; and $h_0(t)$ is proportional to the inverse of the distance to the source (see Equation \ref{eq:GW_strain}). Following \citep{Sowell2019}, for this analysis we set $\Delta T_{\rm SFT}=0.25$\,s (see Section \ref{s2B} for a discussion of $T_{\rm obs}$). 

In the CoCoA stochastic limit (see Section \ref{s2A}), the detection cross-correlation statistic is a Gaussian variable with mean ($\mu_{\rho}$) and standard deviation ($\sigma_{\rho}$) defined as follows (see Equations 4.17 and 4.18 of \cite{Coyne2016}):
\begin{widetext}
\begin{equation}
     \mu_{\rho} = ({\mathcal A}_{+}^2 F^2_{+,H} + {\mathcal A}_{\times}^2 F^2_{\times,H})({\mathcal A}_{+}^2 F^2_{+,L} + {\mathcal A}_{\times}^2 F^2_{\times,L}) \frac{\Delta T^2_{SFT}}{2} \sum_{I}^{N_{SFT}} \frac{h_0^2(T_I)}{S_n^H [f_{k,I}] S_n^L [f_{k,I}] },
 \label{eq:mean_wideeq1}
\end{equation}
\end{widetext}
\begin{widetext}
\begin{equation}
     \sigma_{\rho}^2 = (\mathcal{A}_{+}^2 F^2_{+,H} + \mathcal{A}_{\times}^2 F^2_{\times,H})(\mathcal{A}_{+}^2 F^2_{+,L} + \mathcal{A}_{\times}^2 F^2_{\times,L}) \frac{\Delta T^2_{SFT}}{2} \sum_{I}^{N_{SFT}} \frac{1}{S_n^H [f_{k,I}] S_n^L [f_{k,I}]},
 \label{eq:rho_wideeq2}
\end{equation}
\end{widetext}
where $S_n$ is the single-sided Power Spectral Density (PSD)
of the noise; the $H$ and $L$ subscripts denote quantities calculated for two different detectors, i.e., in this example, LIGO Hanford and LIGO Livingston, respectively.

In determining distance horizons for CoCoA searches, following \citep{Coyne2016} and \citep{Sowell2019}, we set a false alarm probability ($\alpha$) at 0.1\%, a false dismissal probability ($1-\gamma$) at 50\%, and antenna factors to values comparable to what would be reasonable to use for background estimation purposes in a GW170817-like search with the LIGO Hanford and Livingston detectors under the approximation of non-time-varying antenna factors, i.e. $(F_+,F_{\times})_H=(-0.092,-0.91)$ and $(F_+,F_{\times})_L=(0.26,0.79)$ (see \citep{Sowell2019}). In the stochastic limit, assuming white Gaussian noise (so that $S_n$ is independent of frequency), these detection conditions imply $h_{\rm rms} (1/d) \gtrsim  h_{\rm stoch}$ with (see Equation 4.23 in \citep{Coyne2016}):
\begin{equation}
    h_{\rm stoch}=\frac{\sqrt{2}\mathcal{S}^{1/2} \Delta T_{SFT}^{-1/2}N_{SFT}^{-1/4}{(S_n^HS_n^L)}^{1/4}}{[(\mathcal{A}^2_+ F^2_{+,H} + \mathcal{A}^2_\times F^2_{\times,H})(\mathcal{A}^2_+ F^2_{+,L} + \mathcal{A}^2_\times F^2_{\times,L})]^{1/4}},
    \label{eq:h_min_stoch}
\end{equation}
where ${\mathcal S}$ is defined as:
\begin{equation}
    {\mathcal S}={\rm erfc}^{-1}({2\alpha})-{\rm erfc}^{-1}({2\gamma}),
\end{equation}
where ${\rm erfc}(x)$ is the complementary error function of $x$ and ${\rm erfc}^{-1}(x)$ is its inverse. The complementary error function is related to the error function by the relation ${\rm erfc}(x) = 1-{\rm erf}(x)$, where the ${\rm erf}(x)$ is the probability that a random variable that is normally distributed with mean zero and standard deviation $1/\sqrt{2}$ falls in the range $[-x, x]$. Hence, the complementary error function represents the area under the two tails of a zero-mean
Gaussian probability density function with standard deviation $1/\sqrt{2}$. 

In the CoCoA matched filter limit, the detection cross-correlation statistic is a $\chi^2$ with 2 degrees of freedom (d.o.f.) and a non-centrality parameter $\lambda$ given by (see Equation 4.35 in \citep{Coyne2016}): 
\begin{equation}
    \lambda = \sum_I^{N_{\rm SFT}} h_0^2(T_I) \left[\frac{ \Delta T_{\rm SFT} \left(\mathcal{A}^2_+ F^2_{+} + \mathcal{A}^2_\times F^2_{\times}\right)}{S_n}\right].
    \label{eq:lambdasum}
\end{equation}
The above Equation assumes again white Gaussian noise and constant antenna factors. In the case of a network of $N_{\rm det}$ detectors with identical PSDs, the above simplifies to (see also Equation 4.37 in \citep{Coyne2016}):
\begin{equation}
    \lambda = \frac{h_{\rm rms}^2 N_{\rm det} T_{\rm obs} \left(\mathcal{A}^2_+ F^2_{+} + \mathcal{A}^2_\times F^2_{\times}\right)}{S_n}.
    \label{eq:lamba-hrms}
\end{equation}
The detection conditions on the false alarm rate and false dismissal probability imply $\lambda \gtrsim \lambda_{\rm match}
$ with $\lambda_{\rm match}$ satisfying the following relation:
\begin{equation}
 \gamma = {\cal F}_{\rm NC\chi^2, 2}\left[{\cal F}^{-1}_{\rm C\chi^2,2}(1-\alpha),\lambda_{\rm match}\right],
 \label{eq:gamma}
\end{equation}
where ${\cal F}_{\chi^2,2}(x)$ is the cumulative density function of a random variable distributed as a central $\chi^2$ with 2 d.o.f., calculated at $x$. In other words, this is the probability that the random variable will take a value $\le x$. ${\cal F}^{-1}_{\rm C\chi^2,2}$ is the inverse cumulative distribution function of a random variable distributed as a $\chi^2$ with 2 d.o.f.  and ${\cal F}_{\rm NC\chi^2,2}[x,y]$ is the probability that a random variable distributed as a non-central $\chi^2$ with 2 d.o.f. and non-centrality parameter $y$, will take a value $\le x$. Combining Equations \ref{eq:lamba-hrms} and \ref{eq:gamma} above one can derive a condition on the minimum $h_{\rm rms}$ for a detection (a condition that determines the distance horizon for a detection at the chosen false alarm and false dismissal probabilities), namely $h_{\rm rms} (1/d)\ge h_{\rm match}$ with: 
\begin{equation}
h_{\rm match} = \sqrt{\frac{\lambda_{\rm match} S_n}{N_{\rm det}T_{\rm obs} ({\cal A}_+^2F^2_++{\cal A}^2_{\times}F^2_{\times})} }.
\end{equation}

In the CoCoA semi-coherent limit, the detection cross-correlation statistic is a $\chi^2$ with $2N_{\rm coh}$ d.o.f.. As noted in \citep{Coyne2016}, the non-centrality parameter for each coherent segment, given by Equation \ref{eq:lambdasum}, will in general vary from one
semi-coherent chunk to the other due to the time-varying
amplitude of the signal. But, since the
total $\lambda$ for the semi-coherent regime is additive across all coherent segments, the total non-centrality parameter remains unchanged from the matched filter limit. Hence, $h_{\rm semicoh}$ such that $h_{\rm rms} (1/d)\ge h_{\rm semicoh}$---the condition that determines the distance horizon for a detection at the chosen false alarm and false dismissal probabilities---can be derived similarly to the matched filter case as:
\begin{equation}
 \gamma = {\cal F}_{\rm NC\chi^2, 2N_{\rm coh}}\left[{\cal F}^{-1}_{\rm C\chi^2,2N_{\rm coh}}(1-\alpha),\lambda_{\rm semicoh}\right]
\end{equation}
and:
\begin{equation}
h_{\rm semicoh} = \sqrt{\frac{\lambda_{\rm semicoh} S_n}{N_{\rm det}T_{\rm obs} ({\cal A}_+^2F^2_++{\cal A}^2_{\times}F^2_{\times})} }.
\end{equation}

\begin{table}
\centering
\caption{CoCoA single-waveform search distance horizon ranges (in Mpc) across detector networks, indicating the minimum and maximum distances for each detection limit, covering the full parameter space. For simplicity, we use $N_{\text{coh}} = 64$ throughout this paper for the semi-coherent method. Spectral leakage effects as described in the text have been applied to the estimated distances reported here. Of relevance here are current and future detector networks (O4HL-CE4020). We include a network at O2 sensitivity (O2HL) for comparison with previous work.  }
\label{tab:Distance_horizon_ranges}
\begin{tabular}{lccc}
\hline
\multicolumn{4}{c}{Distance Horizon (Min, Max) (Mpc)} \\
\hline
Network & Stochastic & Semi-coh & Matched filtering \\
\hline
O2HL & (18, 22) & (39, 59) & (81, 123)\\
O4HL & (40, 48) & (89, 132) & (186, 276) \\
O5HL & (76, 97) & (173, 261) & (361, 545) \\
A\#HL & (163, 180) & (342, 486) & (716, 1017) \\
CE4020 & (559, 652) &(1207, 1777) &  (2524, 3717)\\
\hline
\hline
\end{tabular}
\end{table}

To cover the parameter space shown in Figure \ref{fig:parameter-space-black-dots}, various gridding schemes can be employed. When using a grid with a constant step size, the last needs to strike a balance  between the need for minimizing the mismatch between a real signal present in the data and the closest template GW waveform in the bank adopted for the search, and minimizing the computational cost of the search itself (which is larger for larger template banks associated with smaller step sizes). 

Building on the work of Coyne et al.\,2016 \cite{Coyne2016}, we choose $\Delta B = 10^{12}$\,G and $\Delta R = 20$\,m to ensure that adjacent templates in the bank have a certain degree of overlap in their time-frequency tracks. This gives us a number of templates equal to:
\begin{eqnarray}
\nonumber \frac{(R_{\rm max}-R_{\rm min)}}{\Delta R}\times \frac{(B_{\rm max}-B_{\rm min})}{\Delta B}\\=101 \times 490 \approx 5\times10^4.
\end{eqnarray}

\section{Results}\label{s4}

\begin{figure*}
    \begin{center}        
\includegraphics[width=\textwidth]{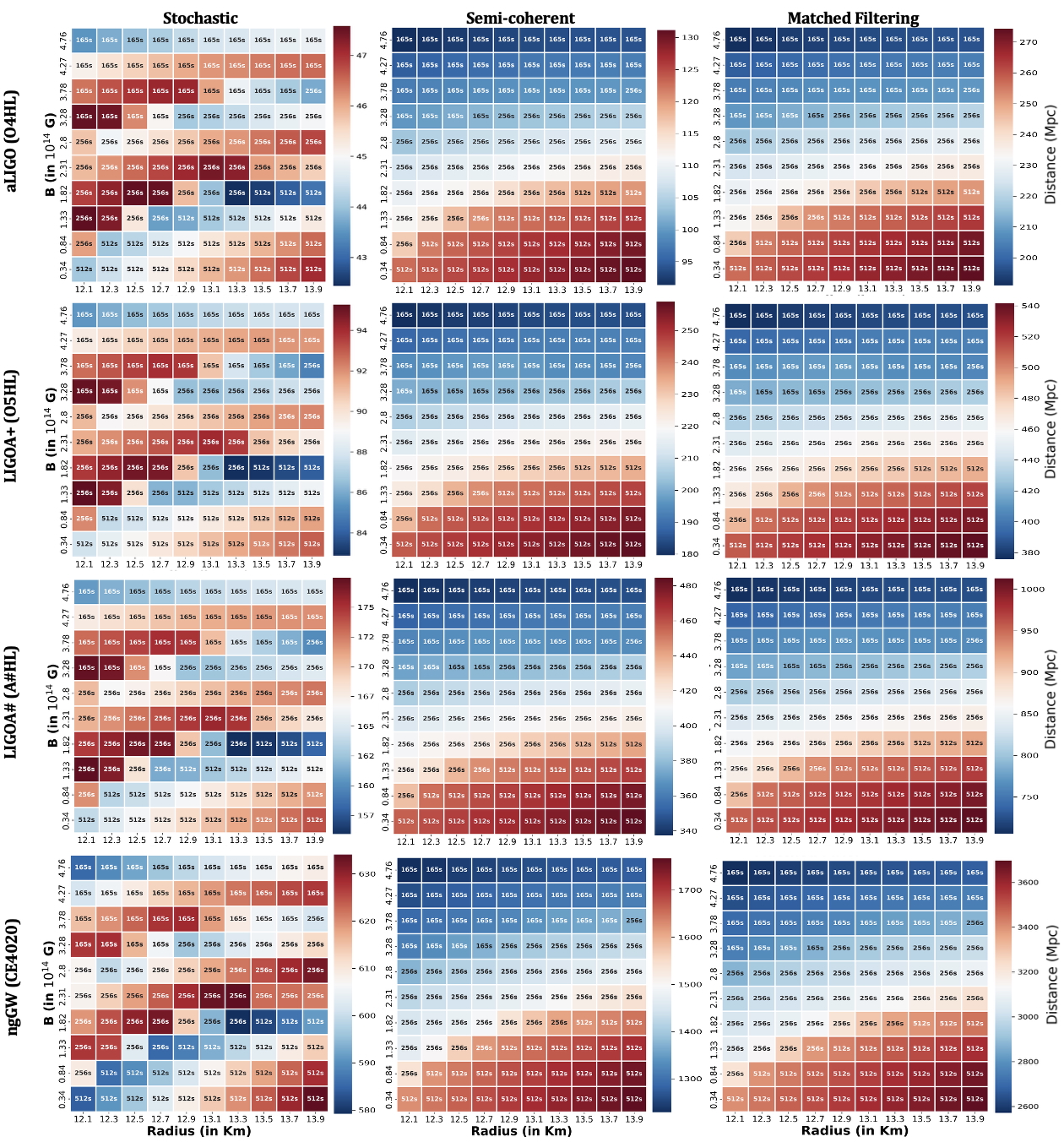}
\end{center}
\caption{Distance horizon estimates for current and future detector networks listed in Table\,\ref{tab:detector-network} using secular bar-mode time-frequency templates grouped in three different observation times ($T_{\rm obs} = 165$; $T_{\rm obs} = 256$\,s; and $T_{\rm obs} = 512$\,s). The horizon distances are shown for the three different regimes of CoCoA: stochastic (left), semi-coherent searches (center), and matched filter (right). In all panels, we color code the distance horizons (in Mpc) reached for the specific values of the magnetic field $B$ (in units of $10^{14}$\,G, vertical axis) and magnetar radius $R$ (in km, horizontal axis), and we report within each $B$-$R$ pixel the corresponding waveform duration (observing time, in seconds). Note that color wedges refer to different distance ranges in the various panels. For reference, GW170817 was located at a distance of 40\,Mpc. See Section\,\ref{s4} for discussion. \label{fig:distance_horizon_estimates}   }
\end{figure*}

In Figure \ref{fig:distance_horizon_estimates} and Table \ref{tab:Distance_horizon_ranges}, we present the distance horizons for single-waveform CoCoA searches assuming a  GW170817-like template bank of secular bar-mode signals. Following \citep{Coyne2016}, we corrected these distances to account for the potential spectral leakage effect of a real search, by multiplying the analytically-derived values by a factor of $\sqrt{0.77}$. We assume the search to be carried with the detector networks listed in Table \ref{tab:detector-network}.

The following trends are evident from Figure \ref{fig:distance_horizon_estimates}. First, for all values of the magnetic field $B$ and magnetar radius $R$,  and all observation times ($165-512$\,s), the stochastic search is the least sensitive (smallest horizon distances) and the matched filter search is the most sensitive (largest horizon distances), as expected. Second, in the stochastic case, there is little dependence of the horizon distance on the specific values of the magnetic field $B$ and magnetar radius $R$ within the considered ranges, in agreement with the fact that the stochastic approach is the most robust against uncertainties in model parameters.  

For the semi-coherent and matched filter cases (center and right panels), we note the following trends. Combinations of the $B$ and $R$ parameter values that result in waveforms of shorter duration are correlated with smaller horizon distances. This is mostly determined by the fact that faster-evolving waveforms are strongly correlated with higher values of $B$ which, in turn, imply that less energy is channeled into GWs. For this same reason, for a fixed value of the magnetar radius $R$, distance horizons decrease with increasing magnetic field $B$. We also note that for a given value of the magnetic field $B$, variations in the magnetar radius $R$ in the considered range affect the horizon distance by less than $\approx 10\%$. On the other hand, for a given value of the magnetar radius $R$, different values of $B$ in the considered range change the distance horizon by up to $\approx 25\%$. Hence, a CoCoA search is expected to be more sensitive to the specific value of the field $B$ compared to the radius $R$, and thus the template bank should adopt a finer grid in $B$ than in $R$ values. 

Overall, an important conclusion from Figure\,\ref{fig:distance_horizon_estimates} is that CoCoA’s O4 distance horizons are estimated to be comparable to that of GW170817 even in CoCoA's most robust stochastic configuration, surpassing the reach of all-sky searches for GW signals of similar durations from a recent analysis of LIGO data collected during the first part of the O4 run \citep{2025arXiv250712282T}. Projections for O5 and next-generation detectors are even more promising. 
We stress, however, that CoCoA searches cannot replace all-sky unmodelled ones. As already noted in \citep{Sowell2019}, CoCoA searches are targeted, and hence assume a known trigger time and sky location, and thus are best suited for GW170817-like post-merger analyses where the merger time and localization are known from the inspiral GW signal and/or EM observations \citep{CM09,Coyne2016,Sowell2019}. Additionally, CoCoA relies on the assumed time–frequency morphology of the signal, requiring a template bank of corresponding time–frequency tracks. Hence, even though CoCoA’s semi-coherent framework does allow for some flexibility with tuning the robustness of the search to the expected level of uncertainty in the templates, it is inherently less robust to signal uncertainties than fully unmodeled methods that make no assumptions about the time–frequency evolution of the GW signal \citep{2025arXiv250712282T}.

\begin{figure}
\centering
\includegraphics[width=\columnwidth]{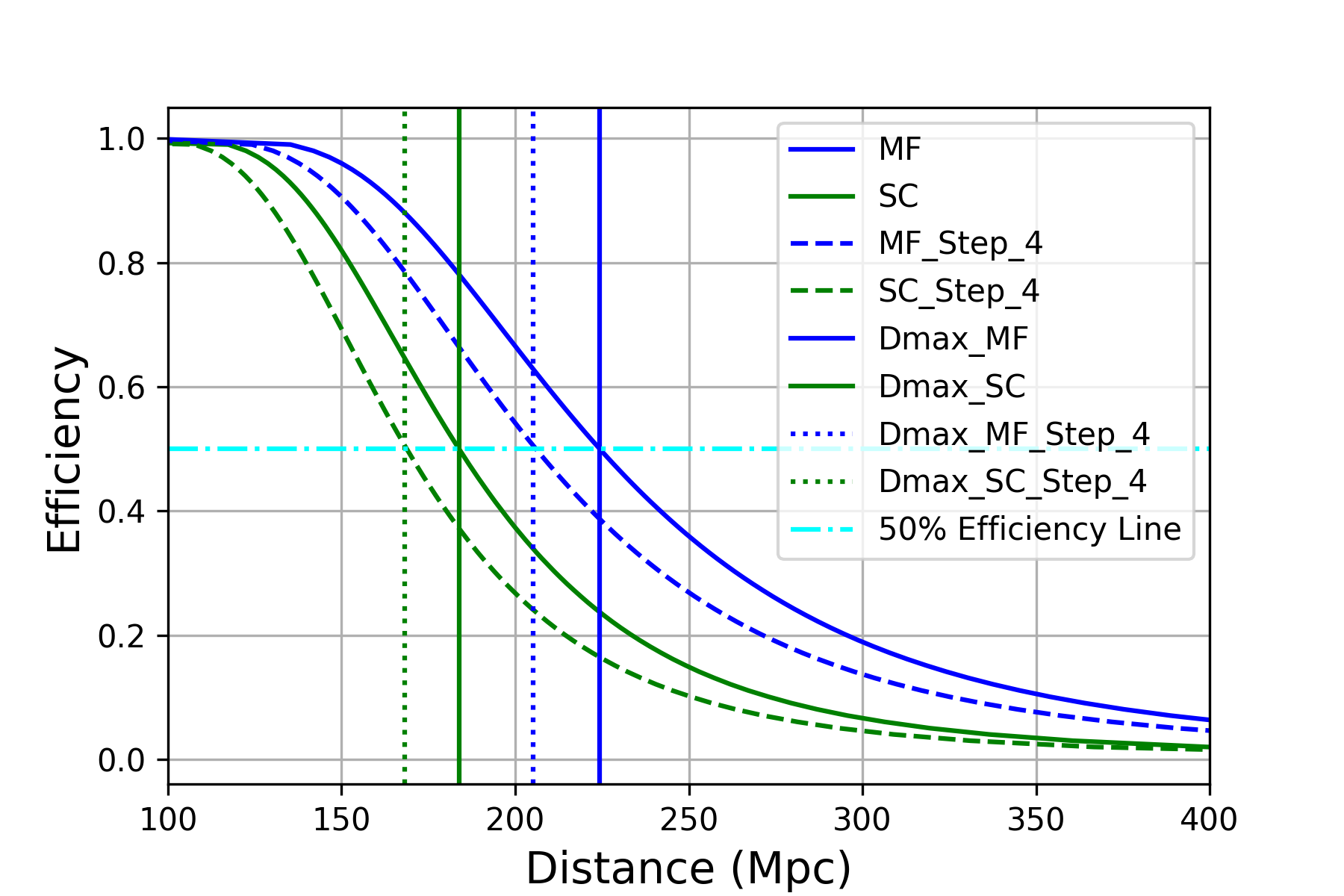}
\caption{Comparison of the match filter (blue) and semi-coherent (green; $N_{coh}$=4) limits of CoCoA. We show the detection efficiency versus distance (in Mpc) for a CoCoA single-waveform search that adopts the ``correct'' time-frequency track (solid) versus a search that adopts a template time-frequency track that is four steps in $\Delta R$ away from the ``correct'' track (dashed).  The horizontal dash-dotted line marks the 50\% detection efficiency. The change in horizon distance (the distance at which one reaches 50\% detection efficiency at a fixed false alarm rate of 0.1\%) when the ``incorrect'' (dashed) time-frequency track is used for the search is larger in the matched filter case (blue) than in the semi-coherent case (green). This highlights the fact that, as expected, a matched filter search is more sensitive but less robust (against signal uncertainties) than a semi-coherent search. \label{fig:det-eff}}
\end{figure}

\section{Comparison with previous results}\label{s4C}
In this Section, we test the accuracy of the results of the Python-based code developed as part of this study by comparing with previous published results for CoCoA. 

First, we calculate distance horizons for the secular bar-mode waveforms CM09long and CM09short used in \citep{Coyne2016}. These are reported in Table \ref{tab:Table-1-CCO}, and are in very good agreement with the results reported in Table 1 of \citeauthor{Coyne2016}.

Then, we repeat the exploration of gridding schemes proposed by \citep{Coyne2016} by calculating the horizon distances that are achieved in CoCoA single-waveform searches when the ``correct'' time-frequency track is CM09long, and the time-frequency track used for the search is mismatched by steps of smaller or larger sizes in the secular bar-mode model parameters $(M,R,B)$. As evident from Tables \ref{tab:Table-2-CCO} and \ref{tab:Table-3-CCO}, also in this case we find our results to be in excellent agreement with those reported in Tables 2 and 3 of \citeauthor{Coyne2016}.
\begin{table}
    \centering
    \begin{tabular}{ccc}
        \hline
        {Waveform} & Matched Filter  & Stochastic  \\
         &  (Mpc) &  (Mpc) \\
        \hline
        CM09long ($\beta = 0.20$) & 104 & 21 \\
        \hline
        CM09short ($\beta = 0.26$) & 218 & 39 \\ 
        \hline
        \hline
    \end{tabular}
    \caption{We reproduce here the calculations of distance horizons for single-waveform CoCoA searches of the secular bar-mode waveforms CM09long and CM09short. These results show excellent agreement with those reported in Table 1 of \citeauthor{Coyne2016}. }
    \label{tab:Table-1-CCO}
\end{table}

\begin{table}
    \centering
    \begin{tabular}{cccc}
        \hline
        {Large steps}& T$_{\rm coh}$ & Semi-coh & Stochastic \\
         &  (s)& (Mpc) & (Mpc) \\
        \hline
        $\Delta B=10^{12}$\,G & 1 & 22 & 21 \\
        \hline
         $\Delta M=5\times10^{-3}$\,M$_{\odot}$ & 2 & 28 & 21 \\ 
        \hline
        $\Delta R=20$\,m & 2 & 28 & 21 \\
        \hline
        $\Delta$All & 0.5 & 18 & 21 \\
        \hline
        \hline
    \end{tabular}
     \caption{We reproduce here the calculations of distance horizons for single-waveform CoCoA searches of the secular bar-mode waveform CM09long carried out with templates that are $\Delta R$ and/or $\Delta B$ and/or $\Delta M$ away from the ``correct'' values. These results show excellent agreement with those reported in Table 2 of \citeauthor{Coyne2016}. }
    \label{tab:Table-2-CCO}
\end{table}

\begin{table}
    \centering
    \begin{tabular}{cccc}
        \hline
        {Small steps}& $T_{\rm coh}$ & Semi-coh & Stochastic \\
        & (s)& (Mpc) & (Mpc) \\
        \hline
        $\Delta B=10^{10}$\,G & 64 & 61 & 21 \\
        \hline
         $\Delta M=5\times10^{-5}$\,M$_{\odot}$ & 256 & 79 & 21 \\ 
        \hline
        $\Delta R=0.2$\,m & 256 & 78 & 21 \\
        \hline
        $\Delta$All & 64 & 61 & 21 \\
        \hline
        \hline
    \end{tabular}
    \caption{We reproduce here the calculations of distance horizons for single-waveform CoCoA searches of the secular bar-mode waveform CM09long carried out with templates that are $\Delta R$ and/or $\Delta B$ and/or $\Delta M$ away from the ``correct'' values. These results show excellent agreement with those reported in Table 3 of \citeauthor{Coyne2016}. \label{tab:Table-3-CCO}}
\end{table}

\begin{figure}
\centering
\includegraphics[width=\columnwidth]{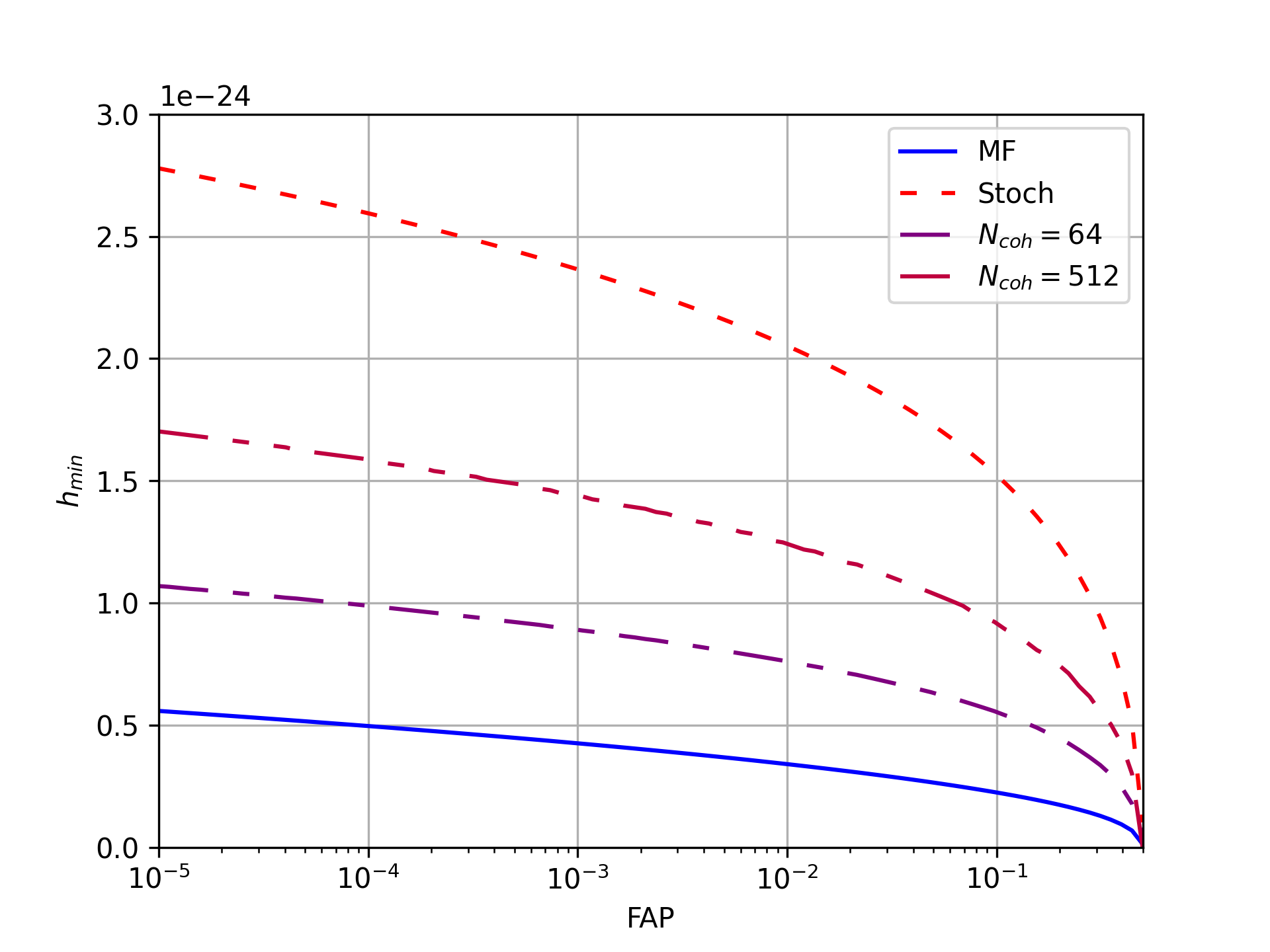}
    \caption{We reproduce Figure 5 in \citeauthor{Coyne2016}, showing the smallest detectable GW amplitude for a CoCoA single-waveform search as a function of false alarm probability $\alpha$ and for a fixed false dismissal probability. See Section\,\ref{s4C} for discussion.}
    \label{fig:Fig-5-CCO}
\end{figure}

Finally, in Figure \ref{fig:Fig-5-CCO} we reproduce the results shown in Figure 5 of \citeauthor{Coyne2016} for the smallest detectable GW amplitude ($h_{\rm min}$) of a CoCoA single-waveform search, plotted versus false alarm probability ($\alpha$) for a set false dismissal probability of $1-\gamma=50\%$. Here, we stress that the following assumptions go into Figure 5, which were not explicitly stated in \citeauthor{Coyne2016}. To reproduce the results, we use the CM09long waveform with an observation duration of \( T_{\text{obs}} = 512 \, \text{s} \), starting at \( t = 295 \, \text{s} \), to evaluate the GW amplitude. The evaluation is performed at the optimal time segment length, \( \Delta T_{\text{SFT}} = 0.25 \, \text{s} \), using a two-detector configuration, which was not explicitly described in the Figure 5 caption of \citeauthor{Coyne2016}.

\section{Summary and outlook}
\label{s5}
In this work, we have developed a Python-based code to streamline the estimation of CoCoA distance horizons for a large set of post-merger waveforms, search approaches (stochastic, semi-coherent, and matched filtering)  and for different networks of ground-based GW detectors. We envision this pipeline to be a useful pre-processing tool for reducing the computational cost of CoCoA searches by identifying, for a given compact binary merger system, the most promising region(s) of the parameter space where a search could focus for a given parameter space gridding scheme.  
In light of the effort that the scientific community is already dedicating to planning for the next generation of ground-based GW detectors, we have carried out our analysis for representative two-detector networks with sensitivities ranging from LIGO O4 to CE. CoCoA post-merger searches are highly promising for reaching astrophysically interesting horizon distances. CoCoA analyses should complement, rather than replace, more robust but less sensitive all-sky searches. It is clear from Table \ref{tab:Distance_horizon_ranges} that the anticipated sensitivity enhancements from upgraded LIGO instruments and next-generation GW detectors will lead to a higher expected detection rate of NS-NS events, underscoring the importance of an optimized approach in post-merger searches.
We note that while these conclusions are drawn under the assumption of Gaussian noise for the GW detectors, \citet{Sowell2019} found relatively good agreement (within 10\%) of the recovered parameters of the CoCoA detection statistic on real data, simulated colored noise, and simulated white Gaussian noise (in both the absence and presence of a signal). Hence, the estimates provided here are robust, at least to first order.

Looking to the future, several improvements are possible in the results here presented, especially for next-generation GW detectors, and as our astrophysical understanding of post-merger scenarios improves. In particular, we have tested the performance of CoCoA only on a specific post-merger model, i.e., that of secularly unstable magnetars, because these are among the most efficient potential sources of post-merger GW signals. However, as discussed e.g. in \citep{2017ApJ...851L..16A,2025arXiv250712282T}, a variety of post-merger scenarios are possible and as predictions for the time-frequency behavior of GW signals associated with these scenarios become more accurate, CoCoA searches could be expanded to hunt for them. In addition, we have carried out tests taking a maximum bar-mode waveform duration of $T_{\rm obs}=512$\,s. However, how long a post-merger remnant can survive before collapsing to a BH is a matter of intense debate. Particularly for the bar-mode scenario, if a post-merger remnant survives longer, its GW signal will have increasing contribution from lower frequencies. While current ground-based GW detectors are not sensitive below $\approx 20$\,Hz, the sensitivity at lower frequencies will improve substantially with next-generation detectors such as CE and the ET. Hence, the need will arise to consider longer waveforms, as well as potentially finer grids in the exploration of allowed parameter spaces.

The biggest progress in our understanding of magnetars as post-merger sources of GWs is likely to come once we are able to collect a larger sample of GW170817-like events and understand the diversity of potential merger outcomes in terms of both their GW and EM emissions. Clearly, the detection of a secular-bar mode GW signal in coincidence with a GRB X-ray afterglow plateau would be an outstanding event. However, even the collection of a larger sample of nearby NS-NS binary systems with total mass less than $\approx 3$\,M$_{\odot}$ and with stringent upper-limits on the most optimistic post-merger scenarios would be a promising route to improve our currently limited knowledge of what happens after two NSs merge. Next generation GW detectors such as CE are projected to be able to detect hundreds to thousands of NS-NS mergers per year within $z\lesssim 0.5$ \citep{2024FrASS..1186748C}. Even if long-lived post-merger remnants are a rare outcome of NS-NS mergers, with such a large sample of nearby events we should be able to probe these rarer outcomes \citep{2025ApJS..280...71P}. Low-mass compact binaries such as the recent O4 candidates S250818k \citep{2025GCN.41437....1L} and S251112cm \citep{2025GCN.41810....1L}, may also produce relatively long-lived ($\gtrsim1$ s) inspiral signals that may fall outside the parameter space targeted by traditional matched-filter searches \cite{2019PhRvL.123p1102A} and could represent another science target for loosely modeled long-duration GW searches.

In summary, it is key to work on optimizing CoCoA and other post-merger searches, to be ready for an era of enhanced detection rates and of unique opportunity for unveiling intermediate-to-long duration post-merger GW signals that would open a new avenue for understanding the physics and astrophysics of NSs.

\begin{acknowledgments}
A.C. and T.K. acknowledge support from the National Science Foundation via the grant PHYS-2438319. R.C. and M.S.P. acknowledge support from the National Science Foundation via the grant PHYS-2512922. The authors are grateful for computational resources provided by the LIGO Laboratory and supported by the National Science Foundation Grants PHY-0757058 and PHY-0823459.
\end{acknowledgments}

\bibliographystyle{plainnat}
\bibliography{References.bib}

@ARTICLE{2022ApJ...926L..31A,
       author = {{Aguilera-Miret}, Ricard and {Vigan{\`o}}, Daniele and {Palenzuela}, Carlos},
        title = "{Universality of the Turbulent Magnetic Field in Hypermassive Neutron Stars Produced by Binary Mergers}",
      journal = {\apjl},
         year = 2022,
        month = feb,
       volume = {926},
       number = {2},
          eid = {L31},
        pages = {L31},
          doi = {10.3847/2041-8213/ac50a7},
archivePrefix = {arXiv},
       eprint = {2112.08406},
 primaryClass = {gr-qc},
       adsurl = {https://ui.adsabs.harvard.edu/abs/2022ApJ...926L..31A}
}

@ARTICLE{2022PhRvD.106b3013P,
       author = {{Palenzuela}, Carlos and {Aguilera-Miret}, Ricard and {Carrasco}, Federico and {Ciolfi}, Riccardo and {Kalinani}, Jay Vijay and {Kastaun}, Wolfgang and {Mi{\~n}ano}, Borja and {Vigan{\`o}}, Daniele},
        title = "{Turbulent magnetic field amplification in binary neutron star mergers}",
      journal = {\prd},
         year = 2022,
        month = jul,
       volume = {106},
       number = {2},
          eid = {023013},
        pages = {023013},
          doi = {10.1103/PhysRevD.106.023013},
archivePrefix = {arXiv},
       eprint = {2112.08413},
 primaryClass = {gr-qc},
       adsurl = {https://ui.adsabs.harvard.edu/abs/2022PhRvD.106b3013P}
}

@ARTICLE{2025ApJS..280...71P,
       author = {{Patel}, Parth and {Corsi}, Alessandra and {Huerta}, E.~A. and {Merfeld}, Kara and {Tiki}, Victoria and {Li}, Zilinghan and {Bicer}, Tekin and {Chard}, Kyle and {Chard}, Ryan and {Foster}, Ian T. and {Gonthier}, Maxime and {Hayot-Sasson}, Valerie and {Nguyen}, Hai Duc and {Pan}, Haochen},
        title = "{Radio Afterglow Detection and AI-driven Response (RADAR): A Federated Framework for Gravitational-wave Event Follow-up}",
      journal = {\apjs},
         year = 2025,
        month = oct,
       volume = {280},
       number = {2},
          eid = {71},
        pages = {71},
          doi = {10.3847/1538-4365/adfbea},
archivePrefix = {arXiv},
       eprint = {2507.14827},
 primaryClass = {astro-ph.HE},
       adsurl = {https://ui.adsabs.harvard.edu/abs/2025ApJS..280...71P}
}

@ARTICLE{2021PhRvD.104j2001A,
       author = {{LIGO Scientific Collaboration} and {Virgo Collaboration} and {KAGRA Collaboration}},
        title = "{All-sky search for long-duration gravitational-wave bursts in the third Advanced LIGO and Advanced Virgo run}",
      journal = {\prd},
         year = 2021,
        month = nov,
       volume = {104},
       number = {10},
          eid = {102001},
        pages = {102001},
          doi = {10.1103/PhysRevD.104.102001},
archivePrefix = {arXiv},
       eprint = {2107.13796},
 primaryClass = {gr-qc},
       adsurl = {https://ui.adsabs.harvard.edu/abs/2021PhRvD.104j2001A}
}

@ARTICLE{2018CQGra..35f5009A,
       author = {{LIGO Scientific Collaboration} and {Vrigo Collaboration}},
        title = "{All-sky search for long-duration gravitational wave transients in the first Advanced LIGO observing run}",
      journal = {Classical and Quantum Gravity},
         year = 2018,
        month = mar,
       volume = {35},
       number = {6},
          eid = {065009},
        pages = {065009},
          doi = {10.1088/1361-6382/aaab76},
archivePrefix = {arXiv},
       eprint = {1711.06843},
 primaryClass = {gr-qc},
       adsurl = {https://ui.adsabs.harvard.edu/abs/2018CQGra..35f5009A}
}

@ARTICLE{2022MNRAS.516.4949S,
       author = {{Sarin}, Nikhil and {Omand}, Conor M.~B. and {Margalit}, Ben and {Jones}, David I.},
        title = "{On the diversity of magnetar-driven kilonovae}",
      journal = {\mnras},
         year = 2022,
        month = nov,
       volume = {516},
       number = {4},
        pages = {4949-4962},
          doi = {10.1093/mnras/stac2609},
archivePrefix = {arXiv},
       eprint = {2205.14159},
 primaryClass = {astro-ph.HE},
       adsurl = {https://ui.adsabs.harvard.edu/abs/2022MNRAS.516.4949S}
}

@ARTICLE{2019ApJ...880L..15M,
       author = {{Margalit}, Ben and {Metzger}, Brian D.},
        title = "{The Multi-messenger Matrix: The Future of Neutron Star Merger Constraints on the Nuclear Equation of State}",
      journal = {\apjl},
         year = 2019,
        month = jul,
       volume = {880},
       number = {1},
          eid = {L15},
        pages = {L15},
          doi = {10.3847/2041-8213/ab2ae2},
archivePrefix = {arXiv},
       eprint = {1904.11995},
 primaryClass = {astro-ph.HE},
       adsurl = {https://ui.adsabs.harvard.edu/abs/2019ApJ...880L..15M}
}

@ARTICLE{2017ApJ...850L..19M,
       author = {{Margalit}, Ben and {Metzger}, Brian D.},
        title = "{Constraining the Maximum Mass of Neutron Stars from Multi-messenger Observations of GW170817}",
      journal = {\apjl},
         year = 2017,
        month = dec,
       volume = {850},
       number = {2},
          eid = {L19},
        pages = {L19},
          doi = {10.3847/2041-8213/aa991c},
archivePrefix = {arXiv},
       eprint = {1710.05938},
 primaryClass = {astro-ph.HE},
       adsurl = {https://ui.adsabs.harvard.edu/abs/2017ApJ...850L..19M}
}

@ARTICLE{2011Natur.478...82N,
       author = {{Nakar}, Ehud and {Piran}, Tsvi},
        title = "{Detectable radio flares following gravitational waves from mergers of binary neutron stars}",
      journal = {\nat},
         year = 2011,
        month = oct,
       volume = {478},
       number = {7367},
        pages = {82-84},
          doi = {10.1038/nature10365},
archivePrefix = {arXiv},
       eprint = {1102.1020},
 primaryClass = {astro-ph.HE},
       adsurl = {https://ui.adsabs.harvard.edu/abs/2011Natur.478...82N}
}

@ARTICLE{Li2024,
       author = {{Li}, Shao-Ze and {Yu}, Yun-Wei and {Gao}, He and {Lan}, Lin},
        title = "{Double Neutron Star Mergers: Are Late-time Radio Signals Overestimated?}",
      journal = {\apj},
         year = 2024,
        month = feb,
       volume = {961},
       number = {2},
          eid = {201},
        pages = {201},
          doi = {10.3847/1538-4357/ad1593},
archivePrefix = {arXiv},
       eprint = {2312.07919},
 primaryClass = {astro-ph.HE},
       adsurl = {https://ui.adsabs.harvard.edu/abs/2024ApJ...961..201L}
}

@ARTICLE{Shan2024,
       author = {{Shan}, Yingze and {Liu}, Xiaoxuan and {Yang}, Xing and {Yuan}, Haoyu and {L{\"u}}, Houjun},
        title = "{GRB 210323A: Signature of Long-lasting Lifetime of Supra-massive Magnetar as the Central Engine from the Merger of Binary Neutron Star}",
      journal = {Research in Astronomy and Astrophysics},
         year = 2024,
        month = aug,
       volume = {24},
       number = {8},
          eid = {085003},
        pages = {085003},
          doi = {10.1088/1674-4527/ad58a7},
       adsurl = {https://ui.adsabs.harvard.edu/abs/2024RAA....24h5003S}
}

@ARTICLE{Gompertz2014,
       author = {{Gompertz}, B.~P. and {O'Brien}, P.~T. and {Wynn}, G.~A.},
        title = "{Magnetar powered GRBs: explaining the extended emission and X-ray plateau of short GRB light curves}",
      journal = {\mnras},
         year = 2014,
        month = feb,
       volume = {438},
       number = {1},
        pages = {240-250},
          doi = {10.1093/mnras/stt2165},
archivePrefix = {arXiv},
       eprint = {1311.1505},
 primaryClass = {astro-ph.HE},
       adsurl = {https://ui.adsabs.harvard.edu/abs/2014MNRAS.438..240G}
}

@misc{ngGWreport,
  howpublished = {\url{https://www.nsf.gov/mps/phy/nggw/mpsac_nggw_subcommittee_report_2024-03-23.pdf}},
  year         = "2024"
}

@misc{PostO5report,
 title={Report of the LSC Post-O5 Study Group},
  howpublished = {\url{https://dcc.ligo.org/LIGO-T2200287/public}},
  year         = "2024"
}

@ARTICLE{2023JCAP...07..068B,
       author = {{Branchesi}, Marica and {Maggiore}, Michele and {Alonso}, David and {Badger}, Charles and {Banerjee}, Biswajit and others},
        title = "{Science with the Einstein Telescope: a comparison of different designs}",
      journal = {\jcap},
         year = 2023,
        month = jul,
       volume = {2023},
       number = {7},
          eid = {068},
        pages = {068},
          doi = {10.1088/1475-7516/2023/07/068},
archivePrefix = {arXiv},
       eprint = {2303.15923},
 primaryClass = {gr-qc},
       adsurl = {https://ui.adsabs.harvard.edu/abs/2023JCAP...07..068B}
}

@ARTICLE{2018LRR....21....3A,
       author = {{LIGO Scientific Collaboration} and {Virgo Collaboration} and {KAGRA Collaboration}},
        title = "{Prospects for observing and localizing gravitational-wave transients with Advanced LIGO, Advanced Virgo and KAGRA}",
      journal = {Living Reviews in Relativity},
         year = 2018,
        month = apr,
       volume = {21},
       number = {1},
          eid = {3},
        pages = {3},
          doi = {10.1007/s41114-018-0012-9},
archivePrefix = {arXiv},
       eprint = {1304.0670},
 primaryClass = {gr-qc},
       adsurl = {https://ui.adsabs.harvard.edu/abs/2018LRR....21....3A}
}

@ARTICLE{2023arXiv230710421G,
       author = {{Gupta}, Ish and {Afle}, Chaitanya and {Arun}, K.~G. and {Bandopadhyay}, Ananya and {Baryakhtar}, Masha and others},
        title = "{Characterizing Gravitational Wave Detector Networks: From A$^\sharp$ to Cosmic Explorer}",
      journal = {arXiv e-prints},
         year = 2023,
        month = jul,
          eid = {arXiv:2307.10421},
        pages = {arXiv:2307.10421},
          doi = {10.48550/arXiv.2307.10421},
archivePrefix = {arXiv},
       eprint = {2307.10421},
 primaryClass = {gr-qc},
       adsurl = {https://ui.adsabs.harvard.edu/abs/2023arXiv230710421G}
}

@ARTICLE{2024FrASS..1186748C,
       author = {{Corsi}, Alessandra and {Barsotti}, Lisa and {Berti}, Emanuele and {Evans}, Matthew and {Gupta}, Ish and others},
        title = "{Multi-messenger astrophysics of black holes and neutron stars as probed by ground-based gravitational wave detectors: from present to future}",
      journal = {Frontiers in Astronomy and Space Sciences},
         year = 2024,
        month = may,
       volume = {11},
          eid = {1386748},
        pages = {1386748},
          doi = {10.3389/fspas.2024.1386748},
archivePrefix = {arXiv},
       eprint = {2402.13445},
 primaryClass = {astro-ph.HE},
       adsurl = {https://ui.adsabs.harvard.edu/abs/2024FrASS..1186748C}
}

@ARTICLE{2017ApJ...848L..12A,
   author = {{LIGO Scientific Collaboration} and {Virgo Collaboration} and others},
    title = "{Multi-messenger Observations of a Binary Neutron Star Merger}",
  journal = {\apjl},
archivePrefix = "arXiv",
   eprint = {1710.05833},
 primaryClass = "astro-ph.HE",
     year = 2017,
    month = oct,
   volume = 848,
      eid = {L12},
    pages = {L12},
      doi = {10.3847/2041-8213/aa91c9},
   adsurl = {http://adsabs.harvard.edu/abs/2017ApJ...848L..12A}
}

@ARTICLE{2022PhRvD.106j2008A,
       author = {{LIGO Scientific Collaboration} and {Virgo Collaboration} and {KAGRA Collaboration}},
        title = "{All-sky search for continuous gravitational waves from isolated neutron stars using Advanced LIGO and Advanced Virgo O3 data}",
      journal = {Physical Review D},
         year = 2022,
        month = nov,
       volume = {106},
       number = {10},
        pages = {102008},
          doi = {10.1103/PhysRevD.106.102008},
       adsurl = {https://ui.adsabs.harvard.edu/abs/2022PhRvD.106j2008A}
}

@ARTICLE{2023_cosmic_explorer_white_paper,
       author = {{Evans}, Matthew and {Corsi}, Alessandra and {Afle}, Chaitanya and {Ananyeva}, Alena and {Arun}, K.~G. and others},
        title = "{Cosmic Explorer: A Submission to the NSF MPSAC ngGW Subcommittee}",
      journal = {arXiv e-prints},
         year = 2023,
        month = jun,
          eid = {arXiv:2306.13745},
        pages = {arXiv:2306.13745},
          doi = {10.48550/arXiv.2306.13745},
archivePrefix = {arXiv},
       eprint = {2306.13745},
 primaryClass = {astro-ph.IM},
       adsurl = {https://ui.adsabs.harvard.edu/abs/2023arXiv230613745E}
}

@ARTICLE{2017ApJ...851L..16A,
       author = {{LIGO Scientific Collaboration} and {Virgo Collaboration} and {KAGRA Collaboration}},
        title = "{Search for Post-merger Gravitational Waves from the Remnant of the Binary Neutron Star Merger GW170817}",
      journal = {\apjl},
         year = 2017,
        month = dec,
       volume = {851},
       number = {1},
          eid = {L16},
        pages = {L16},
          doi = {10.3847/2041-8213/aa9a35},
archivePrefix = {arXiv},
       eprint = {1710.09320},
 primaryClass = {astro-ph.HE},
       adsurl = {https://ui.adsabs.harvard.edu/abs/2017ApJ...851L..16A}
}

@ARTICLE{Andersson_1998A,
       author = {{Andersson}, Nils},
        title = "{A New Class of Unstable Modes of Rotating Relativistic Stars}",
      journal = {\apj},
         year = 1998,
        month = aug,
       volume = {502},
       number = {2},
        pages = {708-713},
          doi = {10.1086/305919},
archivePrefix = {arXiv},
       eprint = {gr-qc/9706075},
 primaryClass = {gr-qc},
       adsurl = {https://ui.adsabs.harvard.edu/abs/1998ApJ...502..708A}
}

@ARTICLE{Bernardini_2012,
    author = {{Bernardini}, M.~G. and {Margutti}, R. and {Mao}, J. and {Zaninoni}, E. and {Chincarini}, G.},
     title = "{The X-ray light curve of gamma-ray bursts: clues to the central engine}",
   journal = {\aap},
      year = 2012,
     month = mar,
    volume = {539},
     pages = {A3},
       doi = {10.1051/0004-6361/201117820},
    adsurl = {https://ui.adsabs.harvard.edu/abs/2012A&A...539A...3B}
}

@ARTICLE{Bonazzola_1996A&A,
       author = {{Bonazzola}, S. and {Gourgoulhon}, E.},
        title = "{Gravitational waves from pulsars: emission by the magnetic-field-induced distortion.}",
      journal = {\aap},
         year = 1996,
        month = aug,
       volume = {312},
        pages = {675-690},
          doi = {10.48550/arXiv.astro-ph/9602107},
archivePrefix = {arXiv},
       eprint = {astro-ph/9602107},
 primaryClass = {astro-ph},
       adsurl = {https://ui.adsabs.harvard.edu/abs/1996A&A...312..675B}
}

@ARTICLE{Coyne2016,
       author = {{Coyne}, Robert and {Corsi}, Alessandra and {Owen}, Benjamin J.},
        title = "{Cross-correlation method for intermediate-duration gravitational wave searches associated with gamma-ray bursts}",
      journal = {\prd},
         year = 2016,
        month = may,
       volume = {93},
       number = {10},
          eid = {104059},
        pages = {104059},
          doi = {10.1103/PhysRevD.93.104059},
archivePrefix = {arXiv},
       eprint = {1512.01301},
 primaryClass = {gr-qc},
       adsurl = {https://ui.adsabs.harvard.edu/abs/2016PhRvD..93j4059C}
}

@ARTICLE{CM09,
       author = {{Corsi}, Alessandra and {M{\'e}sz{\'a}ros}, Peter},
        title = "{Gamma-ray burst afterglow plateaus and gravitational waves}",
      journal = {Classical and Quantum Gravity},
         year = 2009,
        month = oct,
       volume = {26},
       number = {20},
          eid = {204016},
        pages = {204016},
          doi = {10.1088/0264-9381/26/20/204016},
       adsurl = {https://ui.adsabs.harvard.edu/abs/2009CQGra..26t4016C}
}

@ARTICLE{Dhurandhar2009,
       author = {{Dhurandhar}, Sanjeev and {Krishnan}, Badri and {Mukhopadhyay}, Himan and {Whelan}, John T.},
        title = "{Cross-correlation search for periodic gravitational waves}",
      journal = {\prd},
         year = 2008,
        month = apr,
       volume = {77},
       number = {8},
          eid = {082001},
        pages = {082001},
          doi = {10.1103/PhysRevD.77.082001},
archivePrefix = {arXiv},
       eprint = {0712.1578},
 primaryClass = {gr-qc},
       adsurl = {https://ui.adsabs.harvard.edu/abs/2008PhRvD..77h2001D}
}

@ARTICLE{Gompertz_2013,
    author = {{Gompertz}, B.~P. and {O'Brien}, P.~T. and {Wynn}, G.~A. and {Rowlinson}, A.},
     title = "{Can magnetar spin-down power extended emission in some short GRBs?}",
   journal = {\mnras},
      year = 2013,
     month = may,
    volume = {431},
     pages = {1745-1751},
       doi = {10.1093/mnras/stt303},
    adsurl = {https://ui.adsabs.harvard.edu/abs/2013MNRAS.431.1745G}
}

@ARTICLE{Lai_shapiro_1995,
       author = {{Lai}, Dong and {Shapiro}, Stuart L.},
        title = "{Gravitational Radiation from Rapidly Rotating Nascent Neutron Stars}",
      journal = {\apj},
         year = 1995,
        month = mar,
       volume = {442},
        pages = {259},
          doi = {10.1086/175438},
archivePrefix = {arXiv},
       eprint = {astro-ph/9408053},
 primaryClass = {astro-ph},
       adsurl = {https://ui.adsabs.harvard.edu/abs/1995ApJ...442..259L}
}

@ARTICLE{Marek_all_sky_short_2022,
       author = {{Szczepańczyk}, Marek J. and {Salemi}, Francesco and {Bini}, Sophie and {Mishra}, Tanmaya and {Vedovato}, Gabriele and others},
        title = "{Search for gravitational-wave bursts in the third Advanced LIGO-Virgo run with coherent WaveBurst enhanced by Machine Learning}",
      journal = {arXiv e-prints},
         year = 2022,
        month = oct,
          eid = {arXiv:2210.01754},
        pages = {arXiv:2210.01754},
archivePrefix = {arXiv},
       eprint = {2210.01754},
 primaryClass = {gr-qc},
       adsurl = {https://ui.adsabs.harvard.edu/abs/2022arXiv221001754S}
}

@ARTICLE{Metzger2019LRR,
       author = {{Metzger}, Brian D.},
        title = "{Kilonovae}",
      journal = {Living Reviews in Relativity},
         year = 2019,
        month = dec,
       volume = {23},
       number = {1},
          eid = {1},
        pages = {1},
          doi = {10.1007/s41114-019-0024-0},
archivePrefix = {arXiv},
       eprint = {1910.01617},
 primaryClass = {astro-ph.HE},
       adsurl = {https://ui.adsabs.harvard.edu/abs/2019LRR....23....1M}
}

@ARTICLE{Nousek_GRB,
       author = {{Nousek}, J.~A. and {Kouveliotou}, C. and {Grupe}, D. and {Page}, K.~L. and {Granot}, J. and others},
        title = "{Evidence for a Canonical Gamma-Ray Burst Afterglow Light Curve in the Swift XRT Data}",
      journal = {\apj},
         year = 2006,
        month = may,
       volume = {642},
       number = {1},
        pages = {389-400},
          doi = {10.1086/500724},
archivePrefix = {arXiv},
       eprint = {astro-ph/0508332},
 primaryClass = {astro-ph},
       adsurl = {https://ui.adsabs.harvard.edu/abs/2006ApJ...642..389N}
}

@ARTICLE{Ravi_Lasky_2014,
       author = {{Ravi}, Vikram and {Lasky}, Paul D.},
        title = "{The birth of black holes: neutron star collapse times, gamma-ray bursts and fast radio bursts}",
      journal = {\mnras},
         year = 2014,
        month = jul,
       volume = {441},
       number = {3},
        pages = {2433-2439},
          doi = {10.1093/mnras/stu720},
archivePrefix = {arXiv},
       eprint = {1403.6327},
 primaryClass = {astro-ph.HE},
       adsurl = {https://ui.adsabs.harvard.edu/abs/2014MNRAS.441.2433R}
}

@ARTICLE{Rowlinson_2013,
    author = {{Rowlinson}, A. and {O'Brien}, P.~T. and {Metzger}, B.~D. and {Tanvir}, N.~R. and {Levan}, A.~J.},
     title = "{Signatures of magnetar central engines in short GRB light curves}",
   journal = {\mnras},
      year = 2013,
     month = apr,
    volume = {430},
     pages = {1061-1087},
       doi = {10.1093/mnras/sts683},
    adsurl = {https://ui.adsabs.harvard.edu/abs/2013MNRAS.430.1061R}
}

@ARTICLE{Sarin_Lasky_2021,
       author = {{Sarin}, Nikhil and {Lasky}, Paul D.},
        title = "{The evolution of binary neutron star post-merger remnants: a review}",
      journal = {General Relativity and Gravitation},
         year = 2021,
        month = jun,
       volume = {53},
       number = {6},
          eid = {59},
        pages = {59},
          doi = {10.1007/s10714-021-02831-1},
archivePrefix = {arXiv},
       eprint = {2012.08172},
 primaryClass = {astro-ph.HE},
       adsurl = {https://ui.adsabs.harvard.edu/abs/2021GReGr..53...59S}
}

@ARTICLE{starling_2008,
    author = {{Starling}, R.~L.~C. and {O'Brien}, P.~T. and {Willingale}, R. and {Page}, K.~L. and {Osborne}, J.~P. and others},
     title = "{Swift captures the spectrally evolving prompt emission of GRB070616}",
   journal = {\mnras},
      year = 2008,
     month = feb,
    volume = {384},
     pages = {504-514},
       doi = {10.1111/j.1365-2966.2007.12646.x},
    adsurl = {https://ui.adsabs.harvard.edu/abs/2008MNRAS.384..504S}
}

@ARTICLE{Sowell2019,
       author = {{Sowell}, Eric and {Corsi}, Alessandra and {Coyne}, Robert},
        title = "{Multiwaveform cross-correlation search method for intermediate-duration gravitational waves from gamma-ray bursts}",
      journal = {\prd},
         year = 2019,
        month = dec,
       volume = {100},
       number = {12},
          eid = {124041},
        pages = {124041},
          doi = {10.1103/PhysRevD.100.124041},
archivePrefix = {arXiv},
       eprint = {1906.03998},
 primaryClass = {astro-ph.HE},
       adsurl = {https://ui.adsabs.harvard.edu/abs/2019PhRvD.100l4041S}
}

@ARTICLE{Yi_2014,
    author = {{Yi}, S.~X. and {Dai}, Z.~G. and {Wu}, X.~F. and {Wang}, F.~Y.},
     title = "{X-Ray Afterglow Plateaus of Long Gamma-Ray Bursts: Further Evidence for Millisecond Magnetars}",
   journal = {\apjl},
      year = 2014,
     month = sep,
    volume = {792},
     pages = {L8},
       doi = {10.1088/2041-8205/792/1/L8},
    adsurl = {https://ui.adsabs.harvard.edu/abs/2014ApJ...792L...8Y}
}

@ARTICLE{Zhang_2006,
    author = {{Zhang}, B. and {Fan}, Y.~Z. and {Dyks}, J. and {Kobayashi}, S. and {M{\'e}sz{\'a}ros}, P. and others},
     title = "{Physical Processes Shaping Gamma-Ray Burst X-Ray Afterglow Light Curves: Theoretical Implications from the Swift X-Ray Telescope Observations}",
   journal = {\apj},
      year = 2006,
     month = may,
    volume = {642},
    number = {1},
     pages = {354-370},
       doi = {10.1086/500723},
    adsurl = {https://ui.adsabs.harvard.edu/abs/2006ApJ...642..354Z}
}

@article{2017PhRvL.119p1101A,
  title = {GW170817: Observation of Gravitational Waves from a Binary Neutron Star Inspiral},
  author = {{LIGO Scientific Collaboration} and {Virgo Collaboration}},
  collaboration = {LIGO Scientific Collaboration and Virgo Collaboration},
  journal = {Phys. Rev. Lett.},
  volume = {119},
  issue = {16},
  pages = {161101},
  numpages = {18},
  year = {2017},
  month = {Oct},
  publisher = {American Physical Society},
  doi = {10.1103/PhysRevLett.119.161101},
  url = {https://link.aps.org/doi/10.1103/PhysRevLett.119.161101}
}

@ARTICLE{2021PhRvD.104l2004A,
       author = {{LIGO Scientific Collaboration} and {Virgo Collaboration} and {KAGRA Collaboration}},
        title = "{All-sky search for short gravitational-wave bursts in the third Advanced LIGO and Advanced Virgo run}",
      journal = {Physical Review D},
         year = 2021,
        month = dec,
       volume = {104},
       number = {12},
        pages = {122004},
          doi = {10.1103/PhysRevD.104.122004},
       adsurl = {https://ui.adsabs.harvard.edu/abs/2021PhRvD.104l2004A}
}

@ARTICLE{2019ApJ...875..160A,
       author = {{LIGO Scientific Collaboration} and {Virgo Collaboration}},
        title = "{Search for Gravitational Waves from a Long-lived Remnant of the Binary Neutron Star Merger GW170817}",
      journal = {\apj},
         year = 2019,
        month = apr,
       volume = {875},
       number = {2},
          eid = {160},
        pages = {160},
          doi = {10.3847/1538-4357/ab0f3d},
archivePrefix = {arXiv},
       eprint = {1810.02581},
 primaryClass = {gr-qc},
       adsurl = {https://ui.adsabs.harvard.edu/abs/2019ApJ...875..160A}
}

@ARTICLE{2017ApJ...848L..19C,
       author = {{Chornock}, R. and {Berger}, E. and {Kasen}, D. and {Cowperthwaite}, P.~S. and {Nicholl}, M. and {Villar}, V.~A. and {Alexander}, K.~D. and {Blanchard}, P.~K. and {Eftekhari}, T. and {Fong}, W. and {Margutti}, R. and {Williams}, P.~K.~G. and {Annis}, J. and {Brout}, D. and {Brown}, D.~A. and {Chen}, H. -Y. and {Drout}, M.~R. and {Farr}, B. and {Foley}, R.~J. and {Frieman}, J.~A. and {Fryer}, C.~L. and {Herner}, K. and {Holz}, D.~E. and {Kessler}, R. and {Matheson}, T. and {Metzger}, B.~D. and {Quataert}, E. and {Rest}, A. and {Sako}, M. and {Scolnic}, D.~M. and {Smith}, N. and {Soares-Santos}, M.},
        title = "{The Electromagnetic Counterpart of the Binary Neutron Star Merger LIGO/Virgo GW170817. IV. Detection of Near-infrared Signatures of r-process Nucleosynthesis with Gemini-South}",
      journal = {\apjl},
         year = 2017,
        month = oct,
       volume = {848},
       number = {2},
          eid = {L19},
        pages = {L19},
          doi = {10.3847/2041-8213/aa905c},
archivePrefix = {arXiv},
       eprint = {1710.05454},
 primaryClass = {astro-ph.HE},
       adsurl = {https://ui.adsabs.harvard.edu/abs/2017ApJ...848L..19C}
}

@ARTICLE{2017ApJ...848L..17C,
       author = {{Cowperthwaite}, P.~S. and {Berger}, E. and {Villar}, V.~A. and {Metzger}, B.~D. and {Nicholl}, M. and {Chornock}, R. and {Blanchard}, P.~K. and {Fong}, W. and {Margutti}, R. and {Soares-Santos}, M. and {Alexander}, K.~D. and {Allam}, S. and {Annis}, J. and {Brout}, D. and {Brown}, D.~A. and {Butler}, R.~E. and {Chen}, H. -Y. and {Diehl}, H.~T. and {Doctor}, Z. and {Drout}, M.~R. and {Eftekhari}, T. and {Farr}, B. and {Finley}, D.~A. and {Foley}, R.~J. and {Frieman}, J.~A. and {Fryer}, C.~L. and {Garc{\'\i}a-Bellido}, J. and {Gill}, M.~S.~S. and {Guillochon}, J. and {Herner}, K. and {Holz}, D.~E. and {Kasen}, D. and {Kessler}, R. and {Marriner}, J. and {Matheson}, T. and {Neilsen}, E.~H., Jr. and {Quataert}, E. and {Palmese}, A. and {Rest}, A. and {Sako}, M. and {Scolnic}, D.~M. and {Smith}, N. and {Tucker}, D.~L. and {Williams}, P.~K.~G. and {Balbinot}, E. and {Carlin}, J.~L. and {Cook}, E.~R. and {Durret}, F. and {Li}, T.~S. and {Lopes}, P.~A.~A. and {Louren{\c{c}}o}, A.~C.~C. and {Marshall}, J.~L. and {Medina}, G.~E. and {Muir}, J. and {Mu{\~n}oz}, R.~R. and {Sauseda}, M. and {Schlegel}, D.~J. and {Secco}, L.~F. and {Vivas}, A.~K. and {Wester}, W. and {Zenteno}, A. and {Zhang}, Y. and {Abbott}, T.~M.~C. and {Banerji}, M. and {Bechtol}, K. and {Benoit-L{\'e}vy}, A. and {Bertin}, E. and {Buckley-Geer}, E. and {Burke}, D.~L. and {Capozzi}, D. and {Carnero Rosell}, A. and {Carrasco Kind}, M. and {Castander}, F.~J. and {Crocce}, M. and {Cunha}, C.~E. and {D'Andrea}, C.~B. and {da Costa}, L.~N. and {Davis}, C. and {DePoy}, D.~L. and {Desai}, S. and {Dietrich}, J.~P. and {Drlica-Wagner}, A. and {Eifler}, T.~F. and {Evrard}, A.~E. and {Fernandez}, E. and {Flaugher}, B. and {Fosalba}, P. and {Gaztanaga}, E. and {Gerdes}, D.~W. and {Giannantonio}, T. and {Goldstein}, D.~A. and {Gruen}, D. and {Gruendl}, R.~A. and {Gutierrez}, G. and {Honscheid}, K. and {Jain}, B. and {James}, D.~J. and {Jeltema}, T. and {Johnson}, M.~W.~G. and {Johnson}, M.~D. and {Kent}, S. and {Krause}, E. and {Kron}, R. and {Kuehn}, K. and {Nuropatkin}, N. and {Lahav}, O. and {Lima}, M. and {Lin}, H. and {Maia}, M.~A.~G. and {March}, M. and {Martini}, P. and {McMahon}, R.~G. and {Menanteau}, F. and {Miller}, C.~J. and {Miquel}, R. and {Mohr}, J.~J. and {Neilsen}, E. and {Nichol}, R.~C. and {Ogando}, R.~L.~C. and {Plazas}, A.~A. and {Roe}, N. and {Romer}, A.~K. and {Roodman}, A. and {Rykoff}, E.~S. and {Sanchez}, E. and {Scarpine}, V. and {Schindler}, R. and {Schubnell}, M. and {Sevilla-Noarbe}, I. and {Smith}, M. and {Smith}, R.~C. and {Sobreira}, F. and {Suchyta}, E. and {Swanson}, M.~E.~C. and {Tarle}, G. and {Thomas}, D. and {Thomas}, R.~C. and {Troxel}, M.~A. and {Vikram}, V. and {Walker}, A.~R. and {Wechsler}, R.~H. and {Weller}, J. and {Yanny}, B. and {Zuntz}, J.},
        title = "{The Electromagnetic Counterpart of the Binary Neutron Star Merger LIGO/Virgo GW170817. II. UV, Optical, and Near-infrared Light Curves and Comparison to Kilonova Models}",
      journal = {\apjl},
         year = 2017,
        month = oct,
       volume = {848},
       number = {2},
          eid = {L17},
        pages = {L17},
          doi = {10.3847/2041-8213/aa8fc7},
archivePrefix = {arXiv},
       eprint = {1710.05840},
 primaryClass = {astro-ph.HE},
       adsurl = {https://ui.adsabs.harvard.edu/abs/2017ApJ...848L..17C}
}

@ARTICLE{2017Sci...358.1570D,
       author = {{Drout}, M.~R. and {Piro}, A.~L. and {Shappee}, B.~J. and {Kilpatrick}, C.~D. and {Simon}, J.~D. and {Contreras}, C. and {Coulter}, D.~A. and {Foley}, R.~J. and {Siebert}, M.~R. and {Morrell}, N. and {Boutsia}, K. and {Di Mille}, F. and {Holoien}, T.~W. -S. and {Kasen}, D. and {Kollmeier}, J.~A. and {Madore}, B.~F. and {Monson}, A.~J. and {Murguia-Berthier}, A. and {Pan}, Y. -C. and {Prochaska}, J.~X. and {Ramirez-Ruiz}, E. and {Rest}, A. and {Adams}, C. and {Alatalo}, K. and {Ba{\~n}ados}, E. and {Baughman}, J. and {Beers}, T.~C. and {Bernstein}, R.~A. and {Bitsakis}, T. and {Campillay}, A. and {Hansen}, T.~T. and {Higgs}, C.~R. and {Ji}, A.~P. and {Maravelias}, G. and {Marshall}, J.~L. and {Moni Bidin}, C. and {Prieto}, J.~L. and {Rasmussen}, K.~C. and {Rojas-Bravo}, C. and {Strom}, A.~L. and {Ulloa}, N. and {Vargas-Gonz{\'a}lez}, J. and {Wan}, Z. and {Whitten}, D.~D.},
        title = "{Light curves of the neutron star merger GW170817/SSS17a: Implications for r-process nucleosynthesis}",
      journal = {Science},
         year = 2017,
        month = dec,
       volume = {358},
       number = {6370},
        pages = {1570-1574},
          doi = {10.1126/science.aaq0049},
archivePrefix = {arXiv},
       eprint = {1710.05443},
 primaryClass = {astro-ph.HE},
       adsurl = {https://ui.adsabs.harvard.edu/abs/2017Sci...358.1570D}
}

@ARTICLE{2017Sci...358.1565E,
       author = {{Evans}, P.~A. and {Cenko}, S.~B. and {Kennea}, J.~A. and {Emery}, S.~W.~K. and {Kuin}, N.~P.~M. and {Korobkin}, O. and {Wollaeger}, R.~T. and {Fryer}, C.~L. and {Madsen}, K.~K. and {Harrison}, F.~A. and {Xu}, Y. and {Nakar}, E. and {Hotokezaka}, K. and {Lien}, A. and {Campana}, S. and {Oates}, S.~R. and {Troja}, E. and {Breeveld}, A.~A. and {Marshall}, F.~E. and {Barthelmy}, S.~D. and {Beardmore}, A.~P. and {Burrows}, D.~N. and {Cusumano}, G. and {D'A{\`\i}}, A. and {D'Avanzo}, P. and {D'Elia}, V. and {de Pasquale}, M. and {Even}, W.~P. and {Fontes}, C.~J. and {Forster}, K. and {Garcia}, J. and {Giommi}, P. and {Grefenstette}, B. and {Gronwall}, C. and {Hartmann}, D.~H. and {Heida}, M. and {Hungerford}, A.~L. and {Kasliwal}, M.~M. and {Krimm}, H.~A. and {Levan}, A.~J. and {Malesani}, D. and {Melandri}, A. and {Miyasaka}, H. and {Nousek}, J.~A. and {O'Brien}, P.~T. and {Osborne}, J.~P. and {Pagani}, C. and {Page}, K.~L. and {Palmer}, D.~M. and {Perri}, M. and {Pike}, S. and {Racusin}, J.~L. and {Rosswog}, S. and {Siegel}, M.~H. and {Sakamoto}, T. and {Sbarufatti}, B. and {Tagliaferri}, G. and {Tanvir}, N.~R. and {Tohuvavohu}, A.},
        title = "{Swift and NuSTAR observations of GW170817: Detection of a blue kilonova}",
      journal = {Science},
         year = 2017,
        month = dec,
       volume = {358},
       number = {6370},
        pages = {1565-1570},
          doi = {10.1126/science.aap9580},
archivePrefix = {arXiv},
       eprint = {1710.05437},
 primaryClass = {astro-ph.HE},
       adsurl = {https://ui.adsabs.harvard.edu/abs/2017Sci...358.1565E}
}

@ARTICLE{2017Sci...358.1559K,
       author = {{Kasliwal}, M.~M. and {Nakar}, E. and {Singer}, L.~P. and {Kaplan}, D.~L. and {Cook}, D.~O. and {Van Sistine}, A. and {Lau}, R.~M. and {Fremling}, C. and {Gottlieb}, O. and {Jencson}, J.~E. and {Adams}, S.~M. and {Feindt}, U. and {Hotokezaka}, K. and {Ghosh}, S. and {Perley}, D.~A. and {Yu}, P. -C. and {Piran}, T. and {Allison}, J.~R. and {Anupama}, G.~C. and {Balasubramanian}, A. and {Bannister}, K.~W. and {Bally}, J. and {Barnes}, J. and {Barway}, S. and {Bellm}, E. and {Bhalerao}, V. and {Bhattacharya}, D. and {Blagorodnova}, N. and {Bloom}, J.~S. and {Brady}, P.~R. and {Cannella}, C. and {Chatterjee}, D. and {Cenko}, S.~B. and {Cobb}, B.~E. and {Copperwheat}, C. and {Corsi}, A. and {De}, K. and {Dobie}, D. and {Emery}, S.~W.~K. and {Evans}, P.~A. and {Fox}, O.~D. and {Frail}, D.~A. and {Frohmaier}, C. and {Goobar}, A. and {Hallinan}, G. and {Harrison}, F. and {Helou}, G. and {Hinderer}, T. and {Ho}, A.~Y.~Q. and {Horesh}, A. and {Ip}, W. -H. and {Itoh}, R. and {Kasen}, D. and {Kim}, H. and {Kuin}, N.~P.~M. and {Kupfer}, T. and {Lynch}, C. and {Madsen}, K. and {Mazzali}, P.~A. and {Miller}, A.~A. and {Mooley}, K. and {Murphy}, T. and {Ngeow}, C. -C. and {Nichols}, D. and {Nissanke}, S. and {Nugent}, P. and {Ofek}, E.~O. and {Qi}, H. and {Quimby}, R.~M. and {Rosswog}, S. and {Rusu}, F. and {Sadler}, E.~M. and {Schmidt}, P. and {Sollerman}, J. and {Steele}, I. and {Williamson}, A.~R. and {Xu}, Y. and {Yan}, L. and {Yatsu}, Y. and {Zhang}, C. and {Zhao}, W.},
        title = "{Illuminating gravitational waves: A concordant picture of photons from a neutron star merger}",
      journal = {Science},
         year = 2017,
        month = dec,
       volume = {358},
       number = {6370},
        pages = {1559-1565},
          doi = {10.1126/science.aap9455},
archivePrefix = {arXiv},
       eprint = {1710.05436},
 primaryClass = {astro-ph.HE},
       adsurl = {https://ui.adsabs.harvard.edu/abs/2017Sci...358.1559K}
}

@ARTICLE{2017ApJ...848L..18N,
       author = {{Nicholl}, M. and {Berger}, E. and {Kasen}, D. and {Metzger}, B.~D. and {Elias}, J. and {Brice{\~n}o}, C. and {Alexander}, K.~D. and {Blanchard}, P.~K. and {Chornock}, R. and {Cowperthwaite}, P.~S. and {Eftekhari}, T. and {Fong}, W. and {Margutti}, R. and {Villar}, V.~A. and {Williams}, P.~K.~G. and {Brown}, W. and {Annis}, J. and {Bahramian}, A. and {Brout}, D. and {Brown}, D.~A. and {Chen}, H. -Y. and {Clemens}, J.~C. and {Dennihy}, E. and {Dunlap}, B. and {Holz}, D.~E. and {Marchesini}, E. and {Massaro}, F. and {Moskowitz}, N. and {Pelisoli}, I. and {Rest}, A. and {Ricci}, F. and {Sako}, M. and {Soares-Santos}, M. and {Strader}, J.},
        title = "{The Electromagnetic Counterpart of the Binary Neutron Star Merger LIGO/Virgo GW170817. III. Optical and UV Spectra of a Blue Kilonova from Fast Polar Ejecta}",
      journal = {\apjl},
         year = 2017,
        month = oct,
       volume = {848},
       number = {2},
          eid = {L18},
        pages = {L18},
          doi = {10.3847/2041-8213/aa9029},
archivePrefix = {arXiv},
       eprint = {1710.05456},
 primaryClass = {astro-ph.HE},
       adsurl = {https://ui.adsabs.harvard.edu/abs/2017ApJ...848L..18N}
}

@ARTICLE{2017Natur.551...67P,
       author = {{Pian}, E. and {D'Avanzo}, P. and {Benetti}, S. and {Branchesi}, M. and {Brocato}, E. and {Campana}, S. and {Cappellaro}, E. and {Covino}, S. and {D'Elia}, V. and {Fynbo}, J.~P.~U. and {Getman}, F. and {Ghirlanda}, G. and {Ghisellini}, G. and {Grado}, A. and {Greco}, G. and {Hjorth}, J. and {Kouveliotou}, C. and {Levan}, A. and {Limatola}, L. and {Malesani}, D. and {Mazzali}, P.~A. and {Melandri}, A. and {M{\o}ller}, P. and {Nicastro}, L. and {Palazzi}, E. and {Piranomonte}, S. and {Rossi}, A. and {Salafia}, O.~S. and {Selsing}, J. and {Stratta}, G. and {Tanaka}, M. and {Tanvir}, N.~R. and {Tomasella}, L. and {Watson}, D. and {Yang}, S. and {Amati}, L. and {Antonelli}, L.~A. and {Ascenzi}, S. and {Bernardini}, M.~G. and {Bo{\"e}r}, M. and {Bufano}, F. and {Bulgarelli}, A. and {Capaccioli}, M. and {Casella}, P. and {Castro-Tirado}, A.~J. and {Chassande-Mottin}, E. and {Ciolfi}, R. and {Copperwheat}, C.~M. and {Dadina}, M. and {De Cesare}, G. and {di Paola}, A. and {Fan}, Y.~Z. and {Gendre}, B. and {Giuffrida}, G. and {Giunta}, A. and {Hunt}, L.~K. and {Israel}, G.~L. and {Jin}, Z. -P. and {Kasliwal}, M.~M. and {Klose}, S. and {Lisi}, M. and {Longo}, F. and {Maiorano}, E. and {Mapelli}, M. and {Masetti}, N. and {Nava}, L. and {Patricelli}, B. and {Perley}, D. and {Pescalli}, A. and {Piran}, T. and {Possenti}, A. and {Pulone}, L. and {Razzano}, M. and {Salvaterra}, R. and {Schipani}, P. and {Spera}, M. and {Stamerra}, A. and {Stella}, L. and {Tagliaferri}, G. and {Testa}, V. and {Troja}, E. and {Turatto}, M. and {Vergani}, S.~D. and {Vergani}, D.},
        title = "{Spectroscopic identification of r-process nucleosynthesis in a double neutron-star merger}",
      journal = {\nat},
         year = 2017,
        month = nov,
       volume = {551},
       number = {7678},
        pages = {67-70},
          doi = {10.1038/nature24298},
archivePrefix = {arXiv},
       eprint = {1710.05858},
 primaryClass = {astro-ph.HE},
       adsurl = {https://ui.adsabs.harvard.edu/abs/2017Natur.551...67P}
}

@ARTICLE{2017Natur.551...75S,
       author = {{Smartt}, S.~J. and {Chen}, T. -W. and {Jerkstrand}, A. and {Coughlin}, M. and {Kankare}, E. and {Sim}, S.~A. and {Fraser}, M. and {Inserra}, C. and {Maguire}, K. and {Chambers}, K.~C. and {Huber}, M.~E. and {Kr{\"u}hler}, T. and {Leloudas}, G. and {Magee}, M. and {Shingles}, L.~J. and {Smith}, K.~W. and {Young}, D.~R. and {Tonry}, J. and {Kotak}, R. and {Gal-Yam}, A. and {Lyman}, J.~D. and {Homan}, D.~S. and {Agliozzo}, C. and {Anderson}, J.~P. and {Angus}, C.~R. and {Ashall}, C. and {Barbarino}, C. and {Bauer}, F.~E. and {Berton}, M. and {Botticella}, M.~T. and {Bulla}, M. and {Bulger}, J. and {Cannizzaro}, G. and {Cano}, Z. and {Cartier}, R. and {Cikota}, A. and {Clark}, P. and {De Cia}, A. and {Della Valle}, M. and {Denneau}, L. and {Dennefeld}, M. and {Dessart}, L. and {Dimitriadis}, G. and {Elias-Rosa}, N. and {Firth}, R.~E. and {Flewelling}, H. and {Fl{\"o}rs}, A. and {Franckowiak}, A. and {Frohmaier}, C. and {Galbany}, L. and {Gonz{\'a}lez-Gait{\'a}n}, S. and {Greiner}, J. and {Gromadzki}, M. and {Guelbenzu}, A. Nicuesa and {Guti{\'e}rrez}, C.~P. and {Hamanowicz}, A. and {Hanlon}, L. and {Harmanen}, J. and {Heintz}, K.~E. and {Heinze}, A. and {Hernandez}, M. -S. and {Hodgkin}, S.~T. and {Hook}, I.~M. and {Izzo}, L. and {James}, P.~A. and {Jonker}, P.~G. and {Kerzendorf}, W.~E. and {Klose}, S. and {Kostrzewa-Rutkowska}, Z. and {Kowalski}, M. and {Kromer}, M. and {Kuncarayakti}, H. and {Lawrence}, A. and {Lowe}, T.~B. and {Magnier}, E.~A. and {Manulis}, I. and {Martin-Carrillo}, A. and {Mattila}, S. and {McBrien}, O. and {M{\"u}ller}, A. and {Nordin}, J. and {O'Neill}, D. and {Onori}, F. and {Palmerio}, J.~T. and {Pastorello}, A. and {Patat}, F. and {Pignata}, G. and {Podsiadlowski}, Ph. and {Pumo}, M.~L. and {Prentice}, S.~J. and {Rau}, A. and {Razza}, A. and {Rest}, A. and {Reynolds}, T. and {Roy}, R. and {Ruiter}, A.~J. and {Rybicki}, K.~A. and {Salmon}, L. and {Schady}, P. and {Schultz}, A.~S.~B. and {Schweyer}, T. and {Seitenzahl}, I.~R. and {Smith}, M. and {Sollerman}, J. and {Stalder}, B. and {Stubbs}, C.~W. and {Sullivan}, M. and {Szegedi}, H. and {Taddia}, F. and {Taubenberger}, S. and {Terreran}, G. and {van Soelen}, B. and {Vos}, J. and {Wainscoat}, R.~J. and {Walton}, N.~A. and {Waters}, C. and {Weiland}, H. and {Willman}, M. and {Wiseman}, P. and {Wright}, D.~E. and {Wyrzykowski}, {\L}. and {Yaron}, O.},
        title = "{A kilonova as the electromagnetic counterpart to a gravitational-wave source}",
      journal = {\nat},
         year = 2017,
        month = nov,
       volume = {551},
       number = {7678},
        pages = {75-79},
          doi = {10.1038/nature24303},
archivePrefix = {arXiv},
       eprint = {1710.05841},
 primaryClass = {astro-ph.HE},
       adsurl = {https://ui.adsabs.harvard.edu/abs/2017Natur.551...75S}
}

@ARTICLE{2017ApJ...848L..16S,
       author = {{Soares-Santos}, M. and {Holz}, D.~E. and {Annis}, J. and {Chornock}, R. and {Herner}, K. and {Berger}, E. and {Brout}, D. and {Chen}, H. -Y. and {Kessler}, R. and {Sako}, M. and {Allam}, S. and {Tucker}, D.~L. and {Butler}, R.~E. and {Palmese}, A. and {Doctor}, Z. and {Diehl}, H.~T. and {Frieman}, J. and {Yanny}, B. and {Lin}, H. and {Scolnic}, D. and {Cowperthwaite}, P. and {Neilsen}, E. and {Marriner}, J. and {Kuropatkin}, N. and {Hartley}, W.~G. and {Paz-Chinch{\'o}n}, F. and {Alexander}, K.~D. and {Balbinot}, E. and {Blanchard}, P. and {Brown}, D.~A. and {Carlin}, J.~L. and {Conselice}, C. and {Cook}, E.~R. and {Drlica-Wagner}, A. and {Drout}, M.~R. and {Durret}, F. and {Eftekhari}, T. and {Farr}, B. and {Finley}, D.~A. and {Foley}, R.~J. and {Fong}, W. and {Fryer}, C.~L. and {Garc{\'\i}a-Bellido}, J. and {Gill}, M.~S.~S. and {Gruendl}, R.~A. and {Hanna}, C. and {Kasen}, D. and {Li}, T.~S. and {Lopes}, P.~A.~A. and {Louren{\c{c}}o}, A.~C.~C. and {Margutti}, R. and {Marshall}, J.~L. and {Matheson}, T. and {Medina}, G.~E. and {Metzger}, B.~D. and {Mu{\~n}oz}, R.~R. and {Muir}, J. and {Nicholl}, M. and {Quataert}, E. and {Rest}, A. and {Sauseda}, M. and {Schlegel}, D.~J. and {Secco}, L.~F. and {Sobreira}, F. and {Stebbins}, A. and {Villar}, V.~A. and {Vivas}, K. and {Walker}, A.~R. and {Wester}, W. and {Williams}, P.~K.~G. and {Zenteno}, A. and {Zhang}, Y. and {Abbott}, T.~M.~C. and {Abdalla}, F.~B. and {Banerji}, M. and {Bechtol}, K. and {Benoit-L{\'e}vy}, A. and {Bertin}, E. and {Brooks}, D. and {Buckley-Geer}, E. and {Burke}, D.~L. and {Carnero Rosell}, A. and {Carrasco Kind}, M. and {Carretero}, J. and {Castander}, F.~J. and {Crocce}, M. and {Cunha}, C.~E. and {D'Andrea}, C.~B. and {da Costa}, L.~N. and {Davis}, C. and {Desai}, S. and {Dietrich}, J.~P. and {Doel}, P. and {Eifler}, T.~F. and {Fernandez}, E. and {Flaugher}, B. and {Fosalba}, P. and {Gaztanaga}, E. and {Gerdes}, D.~W. and {Giannantonio}, T. and {Goldstein}, D.~A. and {Gruen}, D. and {Gschwend}, J. and {Gutierrez}, G. and {Honscheid}, K. and {Jain}, B. and {James}, D.~J. and {Jeltema}, T. and {Johnson}, M.~W.~G. and {Johnson}, M.~D. and {Kent}, S. and {Krause}, E. and {Kron}, R. and {Kuehn}, K. and {Kuhlmann}, S. and {Lahav}, O. and {Lima}, M. and {Maia}, M.~A.~G. and {March}, M. and {McMahon}, R.~G. and {Menanteau}, F. and {Miquel}, R. and {Mohr}, J.~J. and {Nichol}, R.~C. and {Nord}, B. and {Ogando}, R.~L.~C. and {Petravick}, D. and {Plazas}, A.~A. and {Romer}, A.~K. and {Roodman}, A. and {Rykoff}, E.~S. and {Sanchez}, E. and {Scarpine}, V. and {Schubnell}, M. and {Sevilla-Noarbe}, I. and {Smith}, M. and {Smith}, R.~C. and {Suchyta}, E. and {Swanson}, M.~E.~C. and {Tarle}, G. and {Thomas}, D. and {Thomas}, R.~C. and {Troxel}, M.~A. and {Vikram}, V. and {Wechsler}, R.~H. and {Weller}, J. and {Dark Energy Survey} and {Dark Energy Camera GW-EM Collaboration}},
        title = "{The Electromagnetic Counterpart of the Binary Neutron Star Merger LIGO/Virgo GW170817. I. Discovery of the Optical Counterpart Using the Dark Energy Camera}",
      journal = {\apjl},
         year = 2017,
        month = oct,
       volume = {848},
       number = {2},
          eid = {L16},
        pages = {L16},
          doi = {10.3847/2041-8213/aa9059},
archivePrefix = {arXiv},
       eprint = {1710.05459},
 primaryClass = {astro-ph.HE},
       adsurl = {https://ui.adsabs.harvard.edu/abs/2017ApJ...848L..16S}
}

@ARTICLE{2017ApJ...848L..27T,
       author = {{Tanvir}, N.~R. and {Levan}, A.~J. and {Gonz{\'a}lez-Fern{\'a}ndez}, C. and {Korobkin}, O. and {Mandel}, I. and {Rosswog}, S. and {Hjorth}, J. and {D'Avanzo}, P. and {Fruchter}, A.~S. and {Fryer}, C.~L. and {Kangas}, T. and {Milvang-Jensen}, B. and {Rosetti}, S. and {Steeghs}, D. and {Wollaeger}, R.~T. and {Cano}, Z. and {Copperwheat}, C.~M. and {Covino}, S. and {D'Elia}, V. and {de Ugarte Postigo}, A. and {Evans}, P.~A. and {Even}, W.~P. and {Fairhurst}, S. and {Figuera Jaimes}, R. and {Fontes}, C.~J. and {Fujii}, Y.~I. and {Fynbo}, J.~P.~U. and {Gompertz}, B.~P. and {Greiner}, J. and {Hodosan}, G. and {Irwin}, M.~J. and {Jakobsson}, P. and {J{\o}rgensen}, U.~G. and {Kann}, D.~A. and {Lyman}, J.~D. and {Malesani}, D. and {McMahon}, R.~G. and {Melandri}, A. and {O'Brien}, P.~T. and {Osborne}, J.~P. and {Palazzi}, E. and {Perley}, D.~A. and {Pian}, E. and {Piranomonte}, S. and {Rabus}, M. and {Rol}, E. and {Rowlinson}, A. and {Schulze}, S. and {Sutton}, P. and {Th{\"o}ne}, C.~C. and {Ulaczyk}, K. and {Watson}, D. and {Wiersema}, K. and {Wijers}, R.~A.~M.~J.},
        title = "{The Emergence of a Lanthanide-rich Kilonova Following the Merger of Two Neutron Stars}",
      journal = {\apjl},
         year = 2017,
        month = oct,
       volume = {848},
       number = {2},
          eid = {L27},
        pages = {L27},
          doi = {10.3847/2041-8213/aa90b6},
archivePrefix = {arXiv},
       eprint = {1710.05455},
 primaryClass = {astro-ph.HE},
       adsurl = {https://ui.adsabs.harvard.edu/abs/2017ApJ...848L..27T}
}

@ARTICLE{2017ApJ...848L..24V,
       author = {{Valenti}, Stefano and {Sand}, David J. and {Yang}, Sheng and {Cappellaro}, Enrico and {Tartaglia}, Leonardo and {Corsi}, Alessandra and {Jha}, Saurabh W. and {Reichart}, Daniel E. and {Haislip}, Joshua and {Kouprianov}, Vladimir},
        title = "{The Discovery of the Electromagnetic Counterpart of GW170817: Kilonova AT 2017gfo/DLT17ck}",
      journal = {\apjl},
         year = 2017,
        month = oct,
       volume = {848},
       number = {2},
          eid = {L24},
        pages = {L24},
          doi = {10.3847/2041-8213/aa8edf},
archivePrefix = {arXiv},
       eprint = {1710.05854},
 primaryClass = {astro-ph.HE},
       adsurl = {https://ui.adsabs.harvard.edu/abs/2017ApJ...848L..24V}
}

@ARTICLE{2017ApJ...851L..21V,
       author = {{Villar}, V.~A. and {Guillochon}, J. and {Berger}, E. and {Metzger}, B.~D. and {Cowperthwaite}, P.~S. and {Nicholl}, M. and {Alexander}, K.~D. and {Blanchard}, P.~K. and {Chornock}, R. and {Eftekhari}, T. and {Fong}, W. and {Margutti}, R. and {Williams}, P.~K.~G.},
        title = "{The Combined Ultraviolet, Optical, and Near-infrared Light Curves of the Kilonova Associated with the Binary Neutron Star Merger GW170817: Unified Data Set, Analytic Models, and Physical Implications}",
      journal = {\apjl},
         year = 2017,
        month = dec,
       volume = {851},
       number = {1},
          eid = {L21},
        pages = {L21},
          doi = {10.3847/2041-8213/aa9c84},
archivePrefix = {arXiv},
       eprint = {1710.11576},
 primaryClass = {astro-ph.HE},
       adsurl = {https://ui.adsabs.harvard.edu/abs/2017ApJ...851L..21V}
}

@ARTICLE{2017ApJ...848L..21A,
       author = {{Alexander}, K.~D. and {Berger}, E. and {Fong}, W. and {Williams}, P.~K.~G. and {Guidorzi}, C. and {Margutti}, R. and {Metzger}, B.~D. and {Annis}, J. and {Blanchard}, P.~K. and {Brout}, D. and {Brown}, D.~A. and {Chen}, H. -Y. and {Chornock}, R. and {Cowperthwaite}, P.~S. and {Drout}, M. and {Eftekhari}, T. and {Frieman}, J. and {Holz}, D.~E. and {Nicholl}, M. and {Rest}, A. and {Sako}, M. and {Soares-Santos}, M. and {Villar}, V.~A.},
        title = "{The Electromagnetic Counterpart of the Binary Neutron Star Merger LIGO/Virgo GW170817. VI. Radio Constraints on a Relativistic Jet and Predictions for Late-time Emission from the Kilonova Ejecta}",
      journal = {\apjl},
         year = 2017,
        month = oct,
       volume = {848},
       number = {2},
          eid = {L21},
        pages = {L21},
          doi = {10.3847/2041-8213/aa905d},
archivePrefix = {arXiv},
       eprint = {1710.05457},
 primaryClass = {astro-ph.HE},
       adsurl = {https://ui.adsabs.harvard.edu/abs/2017ApJ...848L..21A}
}

@ARTICLE{2017Sci...358.1579H,
       author = {{Hallinan}, G. and {Corsi}, A. and {Mooley}, K.~P. and {Hotokezaka}, K. and {Nakar}, E. and {Kasliwal}, M.~M. and {Kaplan}, D.~L. and {Frail}, D.~A. and {Myers}, S.~T. and {Murphy}, T. and {De}, K. and {Dobie}, D. and {Allison}, J.~R. and {Bannister}, K.~W. and {Bhalerao}, V. and {Chandra}, P. and {Clarke}, T.~E. and {Giacintucci}, S. and {Ho}, A.~Y.~Q. and {Horesh}, A. and {Kassim}, N.~E. and {Kulkarni}, S.~R. and {Lenc}, E. and {Lockman}, F.~J. and {Lynch}, C. and {Nichols}, D. and {Nissanke}, S. and {Palliyaguru}, N. and {Peters}, W.~M. and {Piran}, T. and {Rana}, J. and {Sadler}, E.~M. and {Singer}, L.~P.},
        title = "{A radio counterpart to a neutron star merger}",
      journal = {Science},
         year = 2017,
        month = dec,
       volume = {358},
       number = {6370},
        pages = {1579-1583},
          doi = {10.1126/science.aap9855},
archivePrefix = {arXiv},
       eprint = {1710.05435},
 primaryClass = {astro-ph.HE},
       adsurl = {https://ui.adsabs.harvard.edu/abs/2017Sci...358.1579H}
}

@ARTICLE{2018Natur.554..207M,
       author = {{Mooley}, K.~P. and {Nakar}, E. and {Hotokezaka}, K. and {Hallinan}, G. and {Corsi}, A. and {Frail}, D.~A. and {Horesh}, A. and {Murphy}, T. and {Lenc}, E. and {Kaplan}, D.~L. and {de}, K. and {Dobie}, D. and {Chandra}, P. and {Deller}, A. and {Gottlieb}, O. and {Kasliwal}, M.~M. and {Kulkarni}, S.~R. and {Myers}, S.~T. and {Nissanke}, S. and {Piran}, T. and {Lynch}, C. and {Bhalerao}, V. and {Bourke}, S. and {Bannister}, K.~W. and {Singer}, L.~P.},
        title = "{A mildly relativistic wide-angle outflow in the neutron-star merger event GW170817}",
      journal = {\nat},
         year = 2018,
        month = feb,
       volume = {554},
       number = {7691},
        pages = {207-210},
          doi = {10.1038/nature25452},
archivePrefix = {arXiv},
       eprint = {1711.11573},
 primaryClass = {astro-ph.HE},
       adsurl = {https://ui.adsabs.harvard.edu/abs/2018Natur.554..207M}
}

@ARTICLE{2025arXiv250622835M,
       author = {{Merfeld}, Kara and {Corsi}, Alessandra},
        title = "{Probing Binary Neutron Star Merger Ejecta and Remnants with Gravitational Wave and Radio Observations}",
      journal = {arXiv e-prints},
         year = 2025,
        month = jun,
          eid = {arXiv:2506.22835},
        pages = {arXiv:2506.22835},
          doi = {10.48550/arXiv.2506.22835},
archivePrefix = {arXiv},
       eprint = {2506.22835},
 primaryClass = {astro-ph.HE},
       adsurl = {https://ui.adsabs.harvard.edu/abs/2025arXiv250622835M}
}

\end{document}